\documentclass[aps,prd,showkeys,nofootinbib,superscriptaddress]{revtex4-2}
\usepackage{graphicx}
\usepackage{enumerate}\usepackage{amsmath,slashed,tensor,amssymb,epsfig,amsthm,bm,graphicx,graphics,slashed} 
\usepackage[caption=false]{subfig}
\usepackage{hyperref}
\usepackage[utf8]{inputenc}
\usepackage[top=1.5cm, bottom=1.5cm, left=2cm, right=2cm]{geometry}
\usepackage{epsfig,color,bigints}
\newcommand{\be}{\begin{equation}}\newcommand{\ee}{\end{equation}}
\newcommand{\bea}{\begin{eqnarray}}\newcommand{\eea}{\end{eqnarray}}
\newcommand{\brr}{\begin{array}}\newcommand{\err}{\end{array}}
\newcommand{\bit}{\begin{itemize}}\newcommand{\eit}{\end{itemize}}
\newcommand{\ben}{\begin{enumerate}}\newcommand{\een}{\end{enumerate}}

\newcommand{\bbm}{\begin{bmatrix}}\newcommand{\ebm}{\end{bmatrix}}
\newcommand{\ba}{\begin{array}}
\newcommand{\ea}{\end{array}}
\newcommand{\G}{\textbf}

\newtheorem{mydef}{Definition}
\newtheorem{Lemma}{Lemma}
\newtheorem{theorem}{Theorem}
\newcommand{\bd}{\begin{mydef}} \newcommand{\ed}{\end{mydef}}
\newcommand{\bthe}{\begin{theorem}} \newcommand{\ethe}{\end{theorem}}
\newcommand{\ble}{\begin{Lemma}} \newcommand{\ele}{\end{Lemma}}

\newcommand{\dr}{\mathrm{d}}

\def\ha{\frac{1}{2}}

\def\intx{\int \! \! \mathrm{d}^3 \textbf{x}}

\def\ph{\varphi}
\def\lan{\langle}
\def\lf{\left}

\def\non{\nonumber}\def\pa{\partial}\def\ran{\rangle}

\def\ri{\right}
\def\ga{\gamma}\def\Ga{\Gamma}
\def\de{\delta}\def\De{\Delta}

\def\la{\lambda}\def\si{\sigma}

\def\1{{_{1}}}\def\2{{_{2}}}

\def\noHe0{:\;\!\!\;\!\!:H_e(0):\;\!\!\;\!\!:}
\def\noHm0{:\;\!\!\;\!\!:H_\mu(0):\;\!\!\;\!\!:}

\def\lan{\langle}
\def\lf{\left}

\def\non{\nonumber}
\def\pa{\partial}\def\ran{\rangle}

\def\ri{\right}

\def\ga{\gamma}
\def\Ga{\Gamma}\def\de{\delta}\def\De{\Delta}

\def\la{\lambda}
\def\si{\sigma}

\def\1{{_{1}}}\def\2{{_{2}}}

\begin{document}
%%%%%%%%%%%%%%%%%%%%%%%%%%%%%%%%%%%%%%%%%%%%%%%%%%%%%%%%%%%%%%%%%%%%%%%%%%%%%%%%%%%%%%%%%%%%
\title{Effective action approach to the dynamical map}
%%%%%%%%%%%%%%%%%%%%%%%%%%%%%%%%%%%%%%%%%%%%%%%%%%%%%%%%%%%%%%%%%%%%%%%%%%%%%%%%%%%%%%%%%%%
\author{M.~Lewicki}
\email{marek.lewicki@fuw.edu.pl}

\affiliation{Faculty of Physics, University of Warsaw ul. Pasteura 5, 02-093 Warsaw, Poland.}

\author{L.~Smaldone}
\email{lsmaldone@unisa.it}

\affiliation{Dipartimento di Fisica, Universit\`a di Salerno, Via Giovanni Paolo II, 132 84084 Fisciano, Italy}
\affiliation{INFN Sezione di Napoli, Gruppo collegato di Salerno, Italy}
\affiliation{Faculty of Physics, University of Warsaw ul. Pasteura 5, 02-093 Warsaw, Poland.}

\begin{abstract}
The dynamical map represents a fundamental concept in quantum field theory, providing a solution of the field equations in the Fock space of asymptotic fields. In this paper, we show how to express the dynamical map of a scalar field in the language of quantum effective action. This grants us new insights into the study of topological defects in quantum field theory, showing a connection between the usual least-action principle and Umezawa's boson transformation method.
\end{abstract}

\maketitle
%%%%%%%%%%%%%%%%%%%%%%%%%%%%%%%%%%%%%%%%%%%%%%%%%%%%%%%%%%%%%%%%%%%%%%%%%%%
\section{Introduction}
\label{sec:Introduction}
 
Quantum field theory (QFT) is a cornerstone of modern physics and it represents the common language of various disciplines such as particle physics, condensed matter physics and cosmology~\cite{itzykson2012quantum,kleinert2016particles}.
In this framework, interacting fields are operator-valued generalized functions acting on a Hilbert space usually defined with respect to a set of \emph{physical fields}. For example, in the \emph{Lehmann--Symanzik--Zimmermann (LSZ)} formulation of relativistic QFT~\cite{schweber2013introduction}, it is realized through the Fock space of asymptotic $in$ or $out$ fields. The action of interacting fields on the Fock space is known once they are expanded in terms of the asymptotic fields. Such expression is known as \emph{dynamical map} or \emph{Haag expansion}~\cite{Haag:1955ev,Umezawa:1982nv, Umezawa:1993yq,Greenberg:1994zu,Blasone2011}.

Apart from the conceptual importance, the study of the properties and the structure of the dynamical map led to important advancements in non-perturbative QFT~\cite{Umezawa:1982nv, Umezawa:1993yq}. For example, it revealed how symmetries of interacting field equations are \emph{dynamically rearranged} at the level of asymptotic (or quasi-particle, in condensed matter) fields, when the system undergoes spontaneous symmetry breaking (SSB) \cite{Umezawa1965DynamicalRO,Matsumoto:1973hg,Matsumoto:1975fi,DeConcini:1976uk,10.1143/PTP.65.315}.

Probably one of the most fascinating developments based on the analysis of the dynamical map properties is the study of the interaction between quanta of fields and classical extended objects emerging from the underlying quantum dynamics. Such research line, firstly developed by H. Umezawa and collaborators, is based on the \emph{boson transformation theorem}, which allows one to find a class of solutions of the QFT equations parametrized by a $c$-number function $f$ whose specific form depends on the extended object under study. Over the years, this method was applied to describe different quantum systems with extended objects such as topological defects in crystals \cite{Wadati1977ASF,Wadati1978FDP,PhysRevB.18.4077}, solitons \cite{MERCALDO1981193,Blasone:2001aj}, vortices in superconductors \cite{LepUme,LepManUme}, strings \cite{TZE197563} and bags \cite{PhysRevD.18.1192} in hadron physics and recently it has been applied in the context of emergent gravity \cite{Iorio2023,Iorio:2023sav}. The case of systems at finite-temperature was also analyzed \cite{MANKA1986533,MANKA199061}.

In this paper, we express the dynamical map in the language of quantum effective action. Our starting point is the path-integral formulation developed in Ref.\cite{Swanson1981}.
%Starting from the path-integral formulation of the dynamical map developed in Ref.\cite{Swanson1981}, in this paper we rewrite it in the language of quantum effective action.
We first define an operator-valued effective action which can be expanded in a power series of asymptotic fields. Next, we express the dynamical map in terms of it. We explicitly show that the usual least-(effective) action principle is equivalent to taking the vacuum expectation value of the dynamical map. In order to exploit the method, we also perform explicit computations in simple (but relevant) examples, namely real scalar field theories with $\ph^3$ and $\ph^4$ potentials and a complex scalar field theory with a $\ph^4$ potential. Then, we show how the boson transformation method can be formulated in our language, unifying that framework with the usual effective action approach to deal with topological defects and discussing the above examples when the systems undergo spontaneous symmetry breaking. 

The paper is organized as follows: in Section \ref{sec2}, we briefly review the definition and the path-integral formulation of the dynamical map. In Section \ref{sec3} we define a modified effective action and we use it to derive a new form of the dynamical map. In Section \ref{s4} we explain how the developed methods can be applied to systems where SSB occurs. Then, in Section \ref{sec4} we show how the boson transformation method works in our framework and we connect such methodology with the usual one based on the least-action principle to describe topological defects. Finally, in Section \ref{sec5} we present conclusions and perspectives. 
%%%%%%%%%%%%%%%%%%%%%%%%%%%%%%%%%%%%%%%%%%%%%%%%%%%%%%%%%%%%%%%%%
\section{The dynamical map} \label{sec2}
Let us consider the scalar-field Lagrange density
\be \label{lu1}
\mathcal{L}(\ph, \pa_\mu \ph)\ = \ \ha \pa_\mu \ph(x) \, \pa^\mu \ph(x) \ - \ \frac{m^2}{2} \, \ph^2(x)\, - \, V(\ph(x)) \, ,
\ee
The Heisenberg field equation is
\be \label{fieldeq}
\lf(\Box+m^2\ri) \ph(x) \ = \ j(x) \, , 
\ee
where $j(x) = j[\ph(x)]= \frac{\pa \mathcal{L}_{int}}{\pa \ph(x)}$, with
\be \label{olan}
\mathcal{L}_0(x) \ = \ \ha \pa_\mu \ph(x) \, \pa^\mu \ph(x) \ - \ \frac{m^2}{2} \, \ph^2(x)\ \, , \qquad \mathcal{L}_{int}(x) \ = \  \, - \, V(\ph(x)) \, .
\ee
The asymptotic $in/out$ fields satisfy the free Klein--Gordon equation
\be \label{asfieldeq}
\lf(\Box+m^2\ri) \ph_{in/out}(x) \ = \ 0 \, .
\ee

In relativistic QFT the Fock space $\mathcal{F}$ is defined in terms of asymptotic fields (e.g. $in$ fields) and as a result, Eq.\eqref{fieldeq} is completely solved once $\ph$ is expanded in terms of $\ph_{in}$ (\emph{Yang--Feldman} equation) \cite{Yang:1950vi,greiner2013field,ryder1996}
\be \label{yfeq}
\lan a|\ph(x)|b\ran \ = \  \, \lan a|\ph_{in}(x)|b\ran \ + \ \int \!\! \dr^4 y \, \Delta_R(x-y) \, \lan a|j(x)|b\ran \, , 
\ee
where  $\De_R(x-y)=[\ph_{in}(x) \, , \, \ph_{in}(y)] \, \Theta(x-y)$ is the retarded propagator, and $|a\ran, |b\ran \in \mathcal{F}$. This relation only holds in a weak-sense \cite{greiner2003quantum} and then the sandwich with the Fock space states will be understood in the following. Moreover, we omitted the usual renormalization constants. Actually, in general, such expression should be thought of in terms of renormalized fields $\ph=Z^{-\ha}\ph_B$, where $B$ indicates the bare quantities. Then, counterterms should be also included in the Lagrangian \eqref{olan}. However, these will not be explicitly written in the following, until one-loop calculations are presented.

Eq.\eqref{yfeq} can be solved iteratively. The solution is a series involving normally ordered products of the asymptotic fields, known as the dynamical map or Haag expansion
\cite{Haag:1955ev,Umezawa:1982nv, Umezawa:1993yq,Greenberg:1994zu,Blasone2011}
\be \label{schdyn}
\ph(x) \ = \  \sum^\infty_{n=1} \, \frac{1}{n!} \, \int \!\! \dr^4 x_1 \, \ldots \, \dr^4 x_n \,  F_n(x;x_1,\ldots,x_n) \, : \ph_{in}(x_1) \, \ldots \, \ph_{in}(x_n): \, .
\ee
The coefficients of this expansion are truncated retarded Green's functions \cite{schweber2013introduction,Swanson1981}:
\be \label{coefficients}
F_n(x;x_1,\ldots,x_n) \ = \  K_{x_1} \ldots \, K_{x_n} \, \lan R[ \ph(x) \, \ph(x_1) \, \ldots \, \ph(x_n)] \ran_c \, ,
\ee
where $K_x \equiv \Box+m^2$ is the Klein--Gordon operator, $\lan \ldots \ran \equiv \lan 0|\ldots |0\ran$ and the subscript $c$ means they are connected. The retarded product is defined as \cite{schweber2013introduction}
\be \label{retprod}
R[ \ph(x) \, \ph(x_1) \, \ldots \, \ph(x_n)] \ \equiv \  -(-i)^n \, \sum_{P} \, \Theta(x_1-x_2) \ldots \Theta(x_{n-1}-x_n)  \,   \lf[\ldots\lf[\ph(x),\ph(x_1)\ri],\ph(x_2)\ri], \ldots, \ph(x_n)] \, , 
\ee
where the sum is over all permutations. Eq.\eqref{schdyn} (and sometimes its counterparts in non-relativistic QFT) is also  briefly indicated as
\be
\ph(x) \ \equiv \ \ph[x; \ph_{in}] \, .
\ee
Let us remark that $\lan \ph \ran=0$. The case of SSB of a continuous symmetry, where $\lan \ph \ran \neq 0$ can be also analyzed with this language. However, it is more delicate and it will be introduced later on, once we have acquired a deeper insight on the topic.

A dynamical map can be also written for other Heisenberg operators. A famous example is given by the LSZ reduction formula~\cite{itzykson2012quantum,ryder1996}:
\be  \label{smat} \lf. \lf.
S \ = \ :\exp\lf[ \int \!\! \dr^4 x \, \lf(\ph_{in}(x) \, K_x \, \frac{\de}{\de J(x)}\ri)\ri] \, \mathcal{Z}[J]\ri|_{J=0}: \ = \ : \mathcal{Z}[J]\ri|_{J(x)=\ph_{in}(x) \, K_x}: \, , 
\ee
where the generating functional of Green's functions is
\be \label{zj}
\mathcal{Z}[J] \ = \ \mathcal{N}\int \! \mathcal{D}\ph \, \exp\lf[i \int \!\! \dr^4 x \lf(\mathcal{L}(x)+J(x) \ph(x)  \ri)\ri] \, .
\ee
The explicit expansion in terms of $\ph_{in}$ is 
\be
S \ = \  \sum^\infty_{n=0} \, \frac{1}{n!} \int \!\! \dr^4 x_1 \, \ldots \, \dr^4 x_n \,  S_n(x_1,\ldots,x_n) \, : \ph_{in}(x_1) \, \ldots \, \ph_{in}(x_n): \, .
\ee
where
\be
S_n(x_1,\ldots,x_n) \ = \  \, K_{x_1} \ldots \, K_{x_n} \, \lan T[\ph(x_1) \, \ldots \, \ph(x_n)] \ran \, .
\ee
Here $T[\ldots]$ is the usual time-ordered product. The \emph{connected part} of the $S$-matrix is also defined \cite{Jevicki:1987ax}
\be \label{csmat} \lf. \lf.
S_c \ = \ :\exp\lf[ \int \!\! \dr^4 x \, \lf(\ph_{in}(x) \, K_x \, \frac{\de}{\de J(x)}\ri)\ri] \, \mathcal{W}[J]\ri|_{J=0}: \ = \ : \mathcal{W}[J]\ri|_{J(x)=\ph_{in}(x) \, K_x}: \, , 
\ee
where the generating functional of connected Green's functions is $\mathcal{W}[J] \equiv -i \log \mathcal{Z}[J]$. 

In analogy with the expressions \eqref{smat},\eqref{csmat}, a useful functional integral form of the dynamical map of $\ph$ was derived in Ref. \cite{Swanson1981}\footnote{Here the operator ordering is not specificed. However, we will always use normal ordering in the various examples.}
\be \label{phifromw}
\ph(x) \ = \  \lf.\frac{\de}{i  \de J(x)} \, \mathcal{W}[J,\ph_{in}] \ri|_{J=0} \, , 
\ee
with $\mathcal{W}[J,\ph_{in}] \equiv -i \ln \mathcal{Z}[J,\ph_{in}]$ and
\bea
\mathcal{Z}[J,\ph_{in}] & \equiv & \int \! \mathcal{D}\ph  \, e^{i \int \!\! \dr^4 x \lf(\mathcal{L}_0(x)+\mathcal{L}_{int}[\ph- \, \ph_{in}](x)+J(x) (\ph(x)-  \, \ph_{in}(x)) \ri)}
\non \\[2mm] \label{zphin}
& = &\exp\lf(i\int \!\! \dr^4 x \, \mathcal{L}_{int}\lf(\frac{\de}{i\de J(x)}\ri)\ri) \, \int \!\!\mathcal{D}\ph  \, e^{i \int \!\! \dr^4 x \lf(\mathcal{L}_0(x)+J(x) (\ph(x)-  \, \ph_{in}(x)) \ri)}\, , 
\eea
is the generating functional of Green's functions where (c-number) function $\ph$ in the interacting term is formally shifted by the operator $- \, \ph_{in}$, and the logarithm only selects contributions from the connected Feynman diagrams. Notice also that boundary conditions of the functional integration in $\mathcal{Z}$ are chosen so it generates retarded Green's functions (see Eq.\eqref{retprod}). In the next section,  we will derive this expression in the language of effective action, in order to make it simpler to treat theories with SSB and topological defects.

%%%%%%%%%%%%%%%%%%%%%%%%%%%%%%%%%%%%%%%%%%%%%%%%%%%%%%
%%%%%%%%%%%%%%%%%%%%%%%%%%%%%%%%%%%%%%%%%%%%%%%%%%%%%%%%%%%%%%%%%%%%%%%%
\section{Effective action form of the dynamical map} \label{sec3}
\subsection{Dynamical map and effective action}
The \emph{quantum effective action} is usually defined as the functional Legendre transform of $\mathcal{W}[J]$
\be \label{sea}
\Ga[\Phi] \ = \ \mathcal{W}[J] \, - \, \int \!\! \dr^4 x \, J(x) \, \Phi(x) \, ,
\ee
with $\Phi \equiv \frac{\de W}{\de J}$. Note that $\Phi=\lan \ph \ran$ when $J=0$. This is the quantum analog of the classical action. In fact, the physical vacuum configurations of a QFT system can be determined by the (quantum) least action principle \cite{itzykson2012quantum,ryder1996,kleinert2016particles}
\be  \label{gaeqgen}
\lf. \frac{\de \Ga[\Phi]}{\de \Phi(y)} \ri|_{J=0} \ = \ 0 \, . 
\ee
In the symmetric case, the only solution is the trivial one $\Phi(J=0) = \lan \ph\ran=0$, while SSB corresponds to the case of (constant) non-trivial solutions. Finally, extended objects as topological defects and solitons are described by non-constant solutions $\Phi(x)$ so that the vacuum loses translation invariance. We now proceed to consider the symmetric case.

It has been noticed \cite{Jevicki:1987ax} that the $S$ matrix can be rewritten in terms of $\Ga$
\be \lf. \lf.
S_c \ = \ \exp\lf[i  \int \!\! \dr^4 x \, \int \!\!\dr^4 y \, \lf(\ph_{in}(x) \, K_x \, G_c(x,y) \frac{\de}{\de \Phi(y)}\ri)\ri] \, \Ga[\Phi]\ri|_{\Phi=0} \ = \ \Ga[\Phi]\ri|_{\Phi(x)=i \int \!\!\dr^4 y \, \lf(\ph_{in}(y) \, K_y \, G^{T}_c(y,x)\ri)} \, . 
\ee
Here
\be
i G_c(x,y) \ \equiv \ \lf. \lf. \frac{\de \Phi(y)}{\de J(x)}\ri|_{J=0} \ = \ \frac{\de^2 \mathcal{W}[J]}{\de J(x)J(y)}\ri|_{J=0} \, ,  
\ee
is the two-point connected time-ordered Green's function of the interacting theory. 
This formula can be used to write down a loop expansion of the $S$-matrix. We thus try to find a similar expression also for $\ph$. From now on, retarded boundary conditions will be understood when we refer to Green's functions.

In analogy with Eq.\eqref{sea}, we can define an \emph{operator valued effective action}
\be
\Ga[\Phi,\ph_{in}] \ \equiv \ \mathcal{W}[J,\ph_{in}] \ - \ \int \!\! \dr^4 x \, J(x) \, \Phi(x) \, .
\ee
Then
\be
\frac{\de \mathcal{W}[\Phi,\ph_{in}]}{\de J(x)} \ = \ \int \!\! \dr^4 y \, \frac{\de \Ga[\Phi,\ph_{in}]}{\de \Phi(y)}\frac{\de \Phi(y)}{\de J(x)}-\Phi(x) - \int \!\! \dr^4 y \, J(y)\frac{\de \Phi(y)}{\de J(x)} \, .
\ee

Computing this expression at $J=0$, the result follows from Eq.\eqref{phifromw}:
\be \label{efdyn}
\ph(x) \ = \ \ - \ \lf. \int \!\! \dr^4 y  \, \frac{\de \Ga[\Phi,\ph_{in}]}{\de \Phi(y)} \,  R(x,y) \ri|_{\Phi=0} \, .
\ee
\emph{This is the dynamical map in the effective action language}. It immediately follows that
\be \label{gengaeq}
\lf. \frac{\de \Ga[\Phi,\ph_{in}]}{\de \Phi(y)}\ri|_{\Phi=0} \ = \ \int \!\! \dr^4 y \, \Ga^{(2)}(x,y) \ph(y) \, ,
\ee
where $\Ga^{(2)}(x,y)$ is the retarded two-point vertex function \cite{PhysRevD.59.065004}, which is the inverse of the two-point retarded connected Green's function $R(x,y)$
\be
\de^4(x-y) \ = \ \int \!\! \dr^4 z  \, R(x,z) \Ga^{(2)}(z,y) \, .
\ee

Eq.\eqref{efdyn} can be also used to characterize the properties of $\Ga[\Phi,\ph_{in}]$. In fact, we can expand it as
\bea \label{gaex}
&& \Ga[\Phi,\ph_{in}] = \\[2mm] \non
 &&  \sum^{\infty}_{n,m=0}  \frac{1}{n!  m!} \int \!\! \dr^4 x_1 \ldots \int \!\! \dr^4 x_m \int \!\! \dr^4 y_1 \ldots \int \!\! \dr^4 y_n \,\ga^{m,n}(x_1,\ldots,x_m;y_1,\ldots,y_n) \, \Phi(x_1) \ldots \Phi(x_m) \, : \ph_{in}(y_1) \ldots \ph_{in}(y_n): \, .
\eea
The vacuum expectation value
\be
\lan \Ga[\Phi,\ph_{in}]\ran \ = \ \sum^{\infty}_{m=0} \frac{1}{m!} \int \!\! \dr^4 x_1 \ldots \int \!\! \dr^4 x_m  \,   \ga^{m,0}(x_1,\ldots,x_m) \, \Phi(x_1) \ldots \Phi(x_m) \,,  
\ee
should be identified with the usual effective action, so that
\be
\ga^{m,0}(x_1,\ldots,x_m) \ = \ \Ga^{(m)}(x_1, \ldots, x_m) \, , 
\ee
are the vertex functions. Then, taking the vacuum expectation value of Eq.\eqref{gengaeq} we recover Eq.\eqref{gaeqgen}, in such a case with only the trivial solution
\be  \label{gaeq}
\lf. \frac{\de \Ga[\Phi]}{\de \Phi(y)}\ri|_{\Phi=\Phi(J=0)} \ = \ 0 \, . 
\ee
It is evident that the dynamical map is more informative than Eq.~\eqref{gaeq}, because it contains all the higher order normal products of $\ph_{in}$. Below we will see how the vacuum expectation value of the dynamical map leads to Eq.~\eqref{gaeqgen} even in the non-trivial cases when $\Phi(J=0) \neq 0$.

Moreover, starting from
\be
\lf. \frac{\de \Ga[\Phi,\ph_{in}]}{\de \Phi(y)} \ri|_{\Phi=0} \ = \ \sum^{\infty}_{n=0}\frac{1}{n!} \int \!\! \dr^4 y_1 \ldots \int \!\! \dr^4 y_n \, \ga^{1,n}(x;y_1,\ldots,y_n)  : \ph_{in}(y_1) \ldots \ph_{in}(y_n): \ ,
\ee
we have (see Eq.~\eqref{coefficients})
\be \label{ga1n}
\ga^{1,n}(y;x_1,\ldots,x_n)   \ = \ -\int \!\! \dr^4 y \, \Ga^{(2)}(y,x) \, F(x;x_1,\ldots,x_n) \, .
\ee

In order to compute $\Ga[\Phi,\ph_{in}]$ in concrete examples, let us notice that Eqs~\eqref{schdyn} and \eqref{phifromw} can be equivalently rewritten similarly as Eq.~\eqref{csmat} \cite{Swanson1981}
\be  \label{smat1} \lf. 
\ph(x) \ = \ :\exp\lf[ -\int \!\! \dr^4 x' \, \lf(\ph_{in}(x') \, K_{x'} \, \frac{\de}{\de J(x')}\ri)\ri] \, \frac{\de}{i \de J(x)} \mathcal{W}[J]\ri|_{J=0}: \, . 
\ee
By using the definition~\eqref{sea} we can rewrite this expression as
\bea \non 
\ph(x)  & = & -\int \!\! \dr^4  z \, R(x,z)\exp\lf[ -i\int \!\! \dr^4 x' \, \int \!\!\dr^4 y \, \lf(\ph_{in}(x') \, K_{x'} \, R(x',y) \frac{\de}{\de \Phi(y)}\ri)\ri] \, \, \frac{\de}{\de \Phi(z)} \Ga[\Phi]\Big|_{\Phi=0} \\[2mm]
 & = &   -\int \!\! \dr^4 z \, R(x,z) \frac{\de}{\de \Phi(z)} \Ga\lf[\Phi(y)-i \int \!\!\dr^4 x' \, \lf(\ph_{in}(x') \, K_{x'} \, R(x',y)\ri)\ri]\Big|_{\Phi=0} \, . 
\eea
This proves that, in order to compute the dynamical map, we can use the prescription
\be \label{genpre}
\Ga[\Phi,\ph_{in}] \ = \ \Ga\lf[\Phi(x)-i \int \!\!\dr^4 y \, \lf(\ph_{in}(y) \, K_y \, R(y,x)\ri)\ri] \, .
\ee
This becomes particularly simple for tree-level calculations, where $R(x,y)=R_0(x,y)$. If we call $\la$ the coupling constant, we have $\lim_{\la \to 0}R_0(x,y)=i \De_R(x-y)$. Then, if we limit at first order calculations in $\la$, we can take $R(x,y)=i \De_R(x-y)$, and  $K_y\De_R(x-y)=\de^4(x-y)$. Then
\be \label{gashi}
\Ga[\Phi,\ph_{in}] \ = \ \Ga[\Phi+\ph_{in}] \, .
\ee
Another way to prove this result is the following. At the tree level, the effective action coincides with the classical action
\be
\Ga[\Phi] \ = \ \mathcal{A}[\Phi] \, .
\ee
In that case, Eq.\eqref{gashi} reads
\be
\Ga[\Phi,\ph_{in}] = \mathcal{A}[\Phi+ \ph_{in}] \, .
\ee
Then, the dynamical map in the effective action form~\eqref{efdyn} gives
\be \label{efdyntree}
\ph(x) \ = \ \   \, \ph_{in}(x)+\lf. \int \!\! \dr^4 y  \, \frac{\pa \mathcal{L}_{int}\lf(\Phi+ \ph_{in}\ri)}{\pa \Phi(y)} \,  R(x,y) \ri|_{\Phi=0} \ = \  \, \ph_{in}(x)+\int \!\! \dr^4 y  \, \frac{\pa \mathcal{L}_{int}\lf( \ph_{in}\ri)}{\pa \ph_{in}(y)} \,  R(x,y)  \, .
\ee
Therefore, at the lowest order in the coupling constant(s) of the theory
\be
K_x \ph(x) \ = \ \frac{\pa \mathcal{L}_{int}\lf( \ph_{in}\ri)}{\pa \ph_{in}(y)} \ \approx \ \frac{\pa \mathcal{L}_{int}\lf( \ph\ri)}{\pa\ph(y)}\, , 
\ee
which is the field equation at the tree level. This proves the condition \eqref{gashi} works.

Thanks to Eq.\eqref{gashi} we can also compute loop corrections. In fact, by employing the usual loop expansion of the effective action
\be
\Ga[\Phi] \ = \ \mathcal{A}[\Phi] \ + \ \ha {\rm Tr} \lf[\log \frac{\de^2 \mathcal{A}[\Phi] }{\de \Phi(x) \Phi(y)}\ri] \ + \ \ldots \ , 
\ee
where the dots stand for two or higher loop contributions. Using the prescription \eqref{genpre} one gets
\be \label{effgain}
\Ga[\Phi,\ph_{in}] \ = \ \mathcal{A}[\Phi(z)-i\int \!\!\dr^4 y \, \lf(\ph_{in}(y) \, K_y \, R(y,z)\ri)] ] \ + \ \ha {\rm Tr} \lf[\log \frac{\de^2 \mathcal{A}[\Phi(z)-i \int \!\!\dr^4 y \, \lf(\ph_{in}(y) \, K_y \, R(y,z)\ri)] }{\de \Phi(x) \Phi(y)}\ri] \ + \ \ldots \ . 
\ee
In computing this expression one should take into account that, as usual, appropriate counterterms must be included in the Lagrangian. Moreover, one should expand the retarded function $R$ at the appropriate order.
%%%%%%%%%%%%%%%%%%%%%%%%%%%%%%%%%%%%%%%%%%%%%%%%%%%%%%%%%
\subsection{Explicit examples}
Let us consider, for example, the case of a free field. There
\be
\mathcal{W}[J,\ph_{in}] \ = \ \int \!\! \dr^4 y \int \!\! \dr^4 x \, J(x) \De_R(x-y) J(y) \, + \, \int \!\! \dr^4 x \, J(x) \, \ph_{in}(x) \, .
\ee
In this case \eqref{gashi} holds so that
\be
\Ga[\Phi,\ph_{in}] \ = \  -\ha  \int \!\! \dr^4 x \, \lf(\Phi(x)+\ph_{in}(x)\ri)^2 K_x \, .
\ee
Therefore, differentiating $\Ga$ we get
\be
\ph(x) \ = \ \int \!\! \dr^4 y \, \ph_{in}(y) \, K_y \, \De_R(x-y) \ = \ \ph_{in}(x) \, ,
\ee
where we used $K_y \, \De_R(x-y)=\de^4(x-y)$.

Consider now the case
$
V(\ph) \ = \ \frac{\la}{4!} \ph^4
$ \cite{Goldstone:1961eq}. Using the prescription \eqref{gashi} to compute $\Ga[\Phi,\ph_{in}]$, we get 
\be
-\lf. \frac{\de \Ga[\Phi,\ph_{in}]}{\de \Phi(y)} \ri|_{\Phi=0} \ = \  \, \ph_{in}(y) \, K_y  \, + \,  \frac{\la \, }{6} \, \ph^3_{in}(y) \, , 
\ee
which leads to the following dynamical map
\be \label{treedyn}
\ph(x) \ = \   \ph_{in}(x) \ + \ \frac{\la \, }{6} \, \int \!\! \dr^4 y \, \De_R(x-y) \, :\ph^3_{in}(y): \, .
\ee
that solves the corresponding field equation at tree-level (order $\la$)
\be
K_x \, \ph(x) \ = \  \frac{\la \, }{6} \,  \ph^3_{in}(x) \approx  \frac{\la}{6} \,  \ph^3(x) \, .
\ee

Another example we present is
$
V(\ph) \ = \ \frac{\la}{3!} \ph^3
$. Using the prescription~\eqref{gashi} we immediately get 
\be
-\lf. \frac{\de \Ga[\Phi,\ph_{in}]}{\de \Phi(y)} \ri|_{\Phi=0} \ = \   \, \ph_{in}(y) \, K_y \, + \, \frac{\la \, }{2} \, \ph^2_{in}(y) \, , 
\ee
which leads to the following tree-level dynamical map:
\be \label{treedyn1}
\ph(x) \ = \   \ph_{in}(x) + \  \frac{\la \, }{2} \, \int \!\! \dr^4 y \, \De_R(x-y) \, :\ph^2_{in}(y): \, .
\ee

It is easy to see the previous results apply to other field theories. The simplest generalization is the case of a complex scalar field. 
There we have
\bea
&& \mathcal{Z}[J,J^*,\ph_{in},\ph^\dag_{in}] \ = \  \\[2mm] \non
&& \int \! \mathcal{D}\ph \mathcal{D}\ph^* \, e^{i \int \!\! \dr^4 x \lf(-\ph^*(x) K_x \ph(x)+\mathcal{L}_{int}[\ph- \, \ph_{in} \, , \, \ph^*- \, \ph^\dag_{in}](x)+J(x) (\ph^*(x)- \, \ph^\dag_{in}(x)) + J (x) (\ph(x)- \, \ph_{in}(x)) \ri)} \non \\[2mm]
& = & e^{i\int \!\! \dr^4 x \, \mathcal{L}_{int}\lf(\frac{\de}{i\de J(x)} \, , \, \frac{\de}{i\de J^*(x)}\ri)} \, \int \! \mathcal{D}\ph \mathcal{D}\ph^* \, e^{i \int \!\! \dr^4 x \lf(-\ph^*(x) K_x \ph(x)+J(x) (\ph^*(x)- \, \ph^\dag_{in}(x)) + J (x) (\ph(x)- \, \ph_{in}(x)) \ri)}\, . 
\eea
As before, we can define $\mathcal{W}[J,J^*,\ph_{in},\ph^\dag_{in}] \equiv -i \log \mathcal{Z}[J,J^*,\ph_{in},\ph^\dag_{in}]$ and 
\be
\Ga[\Phi,\Phi^*,\ph_{in},\ph^\dag_{in}] \ = \ \mathcal{W}[J,J^*,\ph_{in},\ph^\dag_{in}] \ - \ \int \!\! \dr^4 x \, J^*(x) \, \Phi(x) \ - \ \int \!\! \dr^4 x \, J(x) \, \Phi^*(x) \, .
\ee
Now Eq.\eqref{efdyn} reads
\bea \label{efdyncomp1}
\ph(x) & = & \ - \ \lf. \int \!\! \dr^4 y  \, \frac{\de \Ga[\Phi,\Phi^*,\ph_{in},\ph^\dag_{in}]}{\de \Phi^*(y)} \,  R(x,y) \ri|_{\Phi=\Phi^*=0} \, , \\[2mm]
\ph^\dag(x) & = & \ - \ \lf. \int \!\! \dr^4 y  \, \frac{\de \Ga[\Phi,\Phi^*,\ph_{in},\ph^\dag_{in}]}{\de \Phi(y)} \,  R(x,y) \ri|_{\Phi=\Phi^*=0} \, . \label{efdyncomp2}
\eea

For example, let us consider the usual $U(1)$ invariant Goldstone model where $\mathcal{L}_{int}=-\frac{\la}{6} \lf(\ph^\dag \ph\ri)^2$ \cite{Goldstone:1961eq,PhysRev.127.965}. Repeating the same passages as in the real scalar case, one can prove that at tree level $\Ga[\Phi,\Phi^*,\ph_{in},\ph^\dag_{in}] \ = \ \Ga[\Phi+ \ph_{in},\Phi^*+ \ph^\dag_{in}]$. Then
\bea \label{treedync1}
\ph(x) & = &  \ph_{in}(x) \, + \, \frac{\la \, }{3} \, \int \!\! \dr^4 y \, \De_R(x-y) \, :\lf(\ph^\dag_{in}(y)\ph_{in}(y)\ri)\ph_{in}(y): \, , \\[2mm]
\ph^\dag(x) & = &   \ph^\dag_{in}(x) \ + \ \frac{\la \,}{3} \, \int \!\! \dr^4 y \, \De_R(x-y) \, :\lf(\ph^\dag_{in}(y)\ph_{in}(y)\ri)\ph^\dag_{in}(y): \, ,
\eea
which solves the field equations at the order $\la$
\bea 
K_x \ph(x) & = &    \frac{\la }{3} \, \lf(\ph^\dag_{in}(x)\ph_{in}(x)\ri)\ph_{in}(x) \ \approx \   \frac{\la}{3} \, \lf(\ph^\dag(x)\ph(x)\ri)\ph(x) \, , \\[2mm]
K_x \ph^\dag(x) & = &   \frac{\la }{3} \, \lf(\ph^\dag_{in}(x)\ph_{in}(x)\ri)\ph^\dag_{in}(x) \ \approx \   \frac{\la}{3} \, \lf(\ph^\dag(x)\ph(x)\ri)\ph^\dag(x) \, .
\eea

Before closing this subsection, it is interesting to sketch how the present approach could be used to perform calculations beyond the tree level. Let us go back to the case of real $\la \ph^4$ potential. In one or higher loop calculations we cannot omit counterterms anymore \cite{peskin2018introduction}
\be \label{lur}
\mathcal{L}^r(\ph, \pa_\mu \ph)\ = \ \ha \pa_\mu \ph(x) \, \pa^\mu \ph(x) \ - \ \frac{m^2}{2} \, \ph^2(x)\, - \, \frac{\la}{4!} \ph^4(x) + \frac{\de_Z}{2} \pa_\mu \ph(x) \, \pa^\mu \ph(x) \ - \ \frac{\de_m}{2} \, \ph^2(x)\, - \, \frac{\de_\la}{4!} \ph^4(x)\, .
\ee
For slowly varying $\Phi$, the effective action can be expanded as
\be
\Ga[\Phi] \ = \ \int \!\! \dr^4 x \, \lf(\frac{Z(\Phi(x))}{2} \pa_\mu \Phi(x) \pa^\mu \Phi(x) -V_{eff}(\Phi(x)) \ri) \ + \ \ldots \, . 
\ee
$V_{eff}$ is the \emph{effective potential} and the dots stand for higher derivatives terms. In the above expression, we must perform the substitution $ \Phi(x) \to
\Phi(x)-i \int \!\!\dr^4 y \, \lf(\ph_{in}(y) \, K_y \, R(y,x)\ri)$ (see Eq.\eqref{effgain}) in order to get $\Ga[\Phi,\ph_{in}]$. At the tree level, the $V_{eff}(\Phi)=V(\Phi)$ and $Z(\Phi)=1$, and the calculations we have presented above hold. Using the loop expansion one gets
\bea
\Ga(\Phi) \ = \ \ \intx \lf[ \mathcal{L}^r(x)\  + \ \frac{i}{2} {\rm Tr}\log \lf(\Box+m^2+\frac{\la \Phi^2(x)}{2}\ri)\ri]\, . 
\eea
The calculation of the trace for non-constant field configurations and then the evaluation of both $V_{eff}(\Phi)$ and $Z(\Phi)$ are non-trivial tasks (see e.g. Refs. \cite{Cheyette:1985ue,PhysRevLett.57.1199}). The result for the one-loop (renormalized) $V_{eff}$ and $Z$ is \cite{Cheyette:1985ue,buchbinder1992effective}
\bea
V_{eff}[\Phi](x) & = & \frac{\la\Phi^4(x)}{4!} + \frac{\lf(m^2+\frac{\la \Phi^2(x)}{2}\ri)^2}{64 \pi^2}  \lf[\log\lf(\frac{m^2+\frac{\la \Phi^2(x)}{2}}{\si^2}\ri)-\frac{3}{2}\ri] \\[2mm]  
Z[\Phi](x) & = & 1+\frac{\la^2}{96 \pi^2} \frac{\Phi^2(x)}{2 m^2+\la \Phi^2(x)} \, .
\eea
$\si^2$ being the renormalization group parameter. Actually, both these expressions take contributions from all orders truncated diagrams. Here we limit the accuracy to computations at second order in $\la$. Then
\bea
V_{eff}[\Phi](x) & \approx & \frac{\la\Phi^4(x)}{4!} + \frac{\lf(m^2+\frac{\la \Phi^2(x)}{2}\ri)^2}{64 \pi^2}  \lf[\log\lf(\frac{m^2}{\si^2}\ri)-\frac{3}{2}\ri] = V[\Phi](x)+V_1[\Phi](x) \\[2mm]  
Z[\Phi](x) & \approx & 1+\frac{\la^2}{96 \pi^2} \Phi^2(x) = 1+Z_1[\Phi](x)\, .
\eea
Moreover, one can write 
\be
R(x,y) \approx R_0(x,y)+R_1(x,y)
\ee
Where the term $R_1$ is proportional to $\hbar$. Then. Eq.\eqref{efdyn}, at one loop, can be written as
\be \label{efdyn1loop}
\ph(x) \ = \ \ - \ \lf. \int \!\! \dr^4 y  \, \lf( \frac{\de \Ga_0[\Phi,\ph_{in}]}{\de \Phi(y)} \,  R_0(x,y)+ \frac{\de \Ga_0[\Phi,\ph_{in}]}{\de \Phi(y)} \,  R_1(x,y)+ \frac{\de \Ga_1[\Phi,\ph_{in}]}{\de \Phi(y)} \, R_0(x,y) \ri) \ri|_{\Phi=0} \, ,
\ee
where we disregarded the $\hbar^2$ term and where 
\bea
\Ga_0[\Phi,\ph_{in}] & = & \int \!\! \dr^4 x \, \lf(\pa_\mu \lf(\Phi(x)-i\int \!\!\dr^4 y \, \ph_{in}(y) \, K_y \, R(y,x)\ri) \pa^\mu \lf(\Phi(x)-i\int \!\!\dr^4 y \, \ph_{in}(y) \, K_y \, R(y,x)\ri) \ri. \non \\[2mm]
&-& \lf. V\lf[\Phi(x)-i\int \!\!\dr^4 y \, \lf(\ph_{in}(y) \, K_y \, R(y,x)\ri)\ri] \ri) \, , \\[2mm]
\Ga_1[\Phi,\ph_{in}] & = & \int \!\! \dr^4 x \lf(Z_1[\Phi+\ph_{in}](x) \pa_\mu \lf(\Phi(x)+\ph_{in}(x)\ri)\pa^\mu \lf(\Phi(x)+\ph_{in}(x)\ri)-V_1[\Phi+\ph_{in}](x)\ri) \, . 
\eea
We will stop our sketch of the example computation here. The complete calculation we leave for future work.
%%%%%%%%%%%%%%%%%%%%%%%%%%%%%%%%%%%%%%%%%%%%%%%%%%%%%%%%%%%%%
\section{Spontaneous symmetry breaking} \label{s4}
Let us now deal with the case of SSB, where
\be \label{vevcon}
\lan \ph \ran  \ = \ v \ \neq \ 0 \, .
\ee
In order to grasp the main issues, let us go back to the example of the real $\la \ph^4$ theory. We write the Lagrangian~\eqref{lu1} as
\be \label{lu2}
\mathcal{L}(\ph, \pa_\mu \ph)\ = \ \ha \pa_\mu \ph(x) \, \pa^\mu \ph(x) \ - \ \frac{\mu^2}{2} \, \ph^2(x)\, - \,\frac{\la}{4!} \ph^4(x)\, .
\ee
It is well-known that SSB occurs when $\mu^2 <0$. We then perform the canonical transformation
\be \label{shift}
\phi(x) \ \equiv \  \ph(x)-v \, , \qquad \lan \phi(x)\ran \ = \ 0 \, .
\ee
Now the Lagrangian can be split into a free and an interacting part
\bea \label{lu3}
\mathcal{L}_0(x) & = & \ha \pa_\mu \phi(x) \, \pa^\mu \phi(x) \ - \ \frac{m^2}{2} \phi^2(x) \, , \\[2mm]
\mathcal{L}_{int}(x) & = & \frac{\de m^2}{2} \phi^2(x)- \frac{\mu^2}{2} \, (2 \phi(x) v+v^2)\, - \,\frac{\la}{4!} (\phi+v)^4\, .
\eea
with $\de m^2 \ \equiv \ m^2-\mu^2$, $m$ being the physical mass of $in/out$ fields.

We can now employ the previous considerations and write down the operator-valued effective action, so that
\be
-\lf. \frac{\de \Ga[\Phi,\ph_{in}]}{\de \Phi(y)} \ri|_{\Phi=0} \ = \  \, \ph_{in}(y) \, K_y  \, - \, \de m^2  \ph_{in}(y)+\mu^2 v  \, + \,  \frac{\la }{6} \, ( \ph_{in}(y)+v)^3 \, . 
\ee
Then
\be \label{treedynssb}
\ph(x) \ = \ v +   \ph_{in}(x) \ + \  \int \!\! \dr^4 y \, \De_R(x-y) \, :\lf[-\de m^2  \ph_{in}(y)+\mu^2 v  \, + \,  \frac{\la}{6}(\ph_{in}(y)+v)^3\ri]: \ 
\ee
Because of Eq.\eqref{vevcon}, the condition
\be
\mu^2 v  \, + \,  \frac{\la}{6}v^3 \ = \ 0 \, , 
\ee
must be fulfilled, which gives the usual result $v = \sqrt{\frac{-6 \mu^2}{\la}}$. Moreover, the condition $\de m^2 = \frac{\la}{2}v^2 \Rightarrow m^2=-2 \mu^2$ must hold to cancel the spurious linear terms. Then, the final form of the dynamical map is
\be \label{treedynssbfinal}
\ph(x) \ = \ v +   \ph_{in}(x)  \ + \ \frac{\la \,  \, v}{2} \, \int \!\! \dr^4 y \, \De_R(x-y) \, :\ph^2_{in}(y): \ + \ \frac{\la \, }{6} \, \int \!\! \dr^4 y \, \De_R(x-y) \, :\ph^3_{in}(y): \, .
\ee

In summary, in the case when the system undergoes SSB
\begin{enumerate}
\item
One should perform a canonical transformation $\ph \to \phi$ as in Eq.\eqref{shift}, so that $\lan \phi\ran=0$
\item
The mass of asymptotic fields should appear in $\mathcal{L}_0$.
\end{enumerate}
Then, the previously developed effective action techniques can be safely applied. Let us remark that since SSB is a non-perturbative phenomenon, the effective action approach to the dynamical map is more suitable in that case.

Let us now consider the complex scalar Goldstone model
\be \label{lu4}
\mathcal{L}(\ph,\ph^\dag,\pa_\mu \ph, \pa_\mu \ph^\dag)\ = \  \pa_\mu \ph^\dag(x) \, \pa^\mu \ph(x) \ - \ \mu^2 \, \ph^\dag(x)\ph(x)\, - \,\frac{\la}{6} \lf(\ph^\dag(x)\ph(x)\ri)^2\, .
\ee
As in the real case, SSB occurs when $\mu^2 <0$. We then perform a canonical transformation analogous to Eq.\eqref{shift}
\be
\phi(x) \ = \ \ph(x)-\frac{v}{\sqrt{2}} \, ,
\ee
with a real $v=\sqrt{2} \lan \ph \ran$. Moreover, we define
\be
\phi(x) \ = \ \frac{\psi(x)+i \chi(x)}{\sqrt{2}} \, .
\ee
We split the Lagrangian as
\bea \label{lu5}
\mathcal{L}_0(x) & = & \ha \pa_\mu \psi(x) \, \pa^\mu \psi(x) + \ha \pa_\mu \chi(x) \, \pa^\mu \chi(x) -  \frac{m^2_\psi}{2} \psi^2(x) -  \frac{m^2_\chi}{2} \chi^2(x) \, , \\[2mm]
\mathcal{L}_{int}(x) & = & \frac{\de m^2_\psi}{2} \psi^2(x)+\frac{\de m^2_\chi}{2} \chi^2(x)- \frac{\mu^2}{2} \, (2 \psi(x) v+v^2)\\[2mm] \non
& - & \,\frac{\la}{6} \lf(\frac{\psi^4(x)}{4}+\frac{\chi^4(x)}{4}+\frac{v^4}{4}+\frac{3 v^2\psi^2(x)}{2} + \frac{\chi^2(x)\psi^2(x)}{2}+\frac{v^2 \chi^2(x)}{2}+v \chi^2(x) \psi(x)+v \psi^3(x)+v^3 \psi(x)\ri)\, .
\eea
The dynamical maps of $\psi$ and $\chi$ are given by the expressions
\bea \label{efdyncomp3}
\psi(x) & = & \ - \ \lf. \int \!\! \dr^4 y  \, \frac{\de \Ga[\Psi,X,\psi_{in},\chi_{in}]}{\de \Psi(y)} \,  R_\psi(x,y) \ri|_{\Psi=X=0} \, , \\[2mm]
\chi(x) & = & \ - \ \lf. \int \!\! \dr^4 y  \, \frac{\de \Ga[\Psi,X,\psi_{in},\chi_{in}]}{\de X(y)} \,  R_\chi(x,y) \ri|_{\Psi=X=0} \, , \label{efdyncomp4}
\eea
where $ R_\chi(x,y) = \lan R[\chi(x) \chi(y)]\ran_c$, $ R_\psi(x,y) = \lan R[\psi(x) \psi(y)]\ran_c$. As in the previous cases, at tree level $\Ga[\Psi,X,\psi_{in},\chi_{in}]=\Ga[\Psi+ \psi_{in},X+_\chi \chi_{in}]$. Performing the explicit calculations and imposing that spurious constant terms and linear terms in the dynamical maps cancel out one finds $m_\chi=0$, namely $\chi_{in}$ is the Nambu-Goldstone field, $v=\sqrt{\frac{-6\mu^2}{\la}}$, and $m_\psi=-2 \mu^2$, as expected. The dynamical maps read
\bea \label{chidyn}
\chi(x)  & = &  \chi_{in}+\frac{\la}{6} \int \!\! \dr^4 y \, \De^\chi_R(x-y) \, :(\chi_{in}^3(y)+2 v   \chi_{in}(y) \psi_{in}(y)+  \chi_{in}(y) \psi_{in}^2(y)): \, , \\[2mm] \label{psidyn}
\psi(x)  & = &  \psi_{in}+\frac{\la}{6} \int \!\! \dr^4 y \, \De^\psi_R(x-y) \, :\lf[v(3 \psi^2_{in}(y)+ \chi^2_{in}(y))+  \psi_{in}(y) \chi_{in}^2(y)+\psi^3_{in}(y)\ri]: \, .
\eea
Within the approximations employed, these satisfy the field equations
\bea
\Box \chi(x)  & \approx & \frac{\la}{6} (\chi^3(y)+2 v \chi(y) \psi(y)+ \chi(y) \psi(y)) \, , \\[2mm]
(\Box -2 \mu^2)\psi(x)  & \approx & \frac{\la}{6} \lf[v(3 \psi^2(y)+ \chi^2(y))+ \psi(y) \chi^2(y)+\psi^3(y)\ri] \, .
\eea

As a final remark of this section. it is clear that one or higher loop corrections would modify the above result. Exactly as in the usual approach based on the computation of one-loop effective potential~\cite{PhysRevD.7.1888}, here quantum corrections generally modify the conditions for the cancellation of spurious linear and constants terms in the dynamical map.
%%%%%%%%%%%%%%%%%%%%%%%%%%%%%%%%%%%%%%%%%%%%%%%%%%%%%%%%%%%%%%%%%%%%%%%%%
\section{The boson transformation method and topological defects} \label{sec4}
\subsection{Boson transformations and effective action}
It has been proved \cite{Matsumoto:1975rp} that, performing the transformation
\be
\label{btran}
\ph_{in}(x) \ \rightarrow \ \ph^f_{in}(x) \ = \ \ph_{in}(x) \ + \ f(x) \, , \qquad K_x f(x) \ = \ 0 \, , 
\ee
known as the \emph{boson transformation} \cite{LepUme,LepManUme,Matsumoto:1973hg,Matsumoto:1975rp,Blasone:2001aj}, in the dynamical map, we get a class of solutions of the field equations parametrized by $f$. For example, in the real scalar case 
\be
\ph^f(x) \ \equiv \ \ph[x; \ph_{in}+f] \, ,
\ee
is an \emph{exact} solution of the original field equation \eqref{fieldeq}:
\be \label{fieldeqf}
K_x \ph^f(x) \ = \ j[\ph^f(x)] \, .
\ee

The generator of the boson transformation \eqref{btran} reads\footnote{Note that $Q$ is time independent:
\bea
\dot{Q}(t) & = & \intx \, \nabla^2 f(x) \, \ph_{in}(x)- f(x) \, \nabla^2 \ph_{in}(x)   \\[2mm] \non
& = &  \intx \, \nabla \cdot \lf(\nabla f(x) \, \ph_{in}(x)- f(x) \, \nabla \ph_{in}(x)\ri) \ = \ \int \! \! \dr \G S  \cdot \lf(\nabla f(x) \, \ph_{in}(x)- f(x) \, \nabla \ph_{in}(x)\ri) \ = \ 0 \, , 
\eea
where the surface integral is extended over a surface at an infinite distance. } \cite{Ezawa:1975ua}
\be\label{charge}
Q \  \ = \ \intx \, f(x) \, \overleftrightarrow{\pa_0} \, \ph_{in}(x)  \, ,
\ee
so that

\be \label{bchf}
U(f) \, \ph_{in}(x) \, U^\dag(f) \ = \ \ph_{in}(x) \, + \, f(x) \, , \quad U(f) \ \equiv \ e^{-i  \, Q} \, .
\ee
The solutions $\ph^f$ describe the quantum field $\ph$ affected by the presence of a classical extended object, described by the $c$-number function $f$. Physically, the boson transformation is an inhomogeneous boson condensation on the vacuum of the asymptotic field $\ph_{in}$
\be
\lan \ph^f_{in}(x) \ran \ = \  \lan 0(f)|\ph_{in}(x)|0(f)\ran \ \equiv \ \lan U(f) \, \ph_{in}(x) \, U^\dag(f)\ran \ = \ f(x) \, ,
\ee
so that the vacuum loses translation invariance, due to the presence of the extended object
\be
e^{i u_a P^a} \, |0(f)\ran \ = \ e^{i \intx \, f(x+u) \, \overleftrightarrow{\pa_0} \, \ph_{in}(x)}|0(f)\ran \ \neq \ |0(f)\ran \, .
\ee

The cases when $f$ is not Fourier transformable are particularly relevant. For example, soliton solutions are described by $f$s which diverge at $x \to \pm \infty$ (in 1+1D)~\cite{MERCALDO1981193,Blasone:2001aj}, while topological defects are defined by the condition~\cite{TZE197563}
\be \label{gmunu}
\varepsilon_{\mu \nu \rho \si}j^{\rho \si}(x) \ \equiv \ \lf[\pa_\mu,\pa_\nu\ri] \, f(x) \ \neq \ 0 \, ,
\ee
where $\varepsilon_{\mu \nu \rho \si}$ is the antisymmetric Levi-Civita symbol.
The spacetime evolution of the defects is described on a sub-manifold parameterised by the coordinates $(\si_0,\si_1,\si_2,\si_3)$. For example, in the case of strings, we have
\be \label{jmunu}
j^{\mu \nu}(x) \ = \ \int \!\! \dr^2 \si\,  \, \varepsilon^{a b} \, \frac{\pa y^\mu}{\pa \si^a}\frac{\pa y^\nu}{\pa \si^b} \, \de^4(x-y(\tau,\si)) \, , 
\ee
i.e. the string dynamics is characterized by the function $y$ on the world-sheet with coordinates $(\tau,\si) \equiv (\si_0,\si_1)$~\cite{Matsumoto:1975rp,PhysRevD.31.3052,Sakellariadou1990,Battye:1993jv,vilenkin1994cosmic}. From a formal point of view, the condition~\eqref{gmunu} indicates a non-trivial cohomology group of the spacetime manifold~\cite{nash1988topology}.

Let us apply the boson transformation into the dynamical map \eqref{efdyn}
\be \label{efdynf}
\ph^f(x) \ = \ \ - \ \lf. \int \!\! \dr^4 y  \, \frac{\de \Ga[\Phi,\ph_{in}+f]}{\de \Phi(y)} \,  R(x,y) \ri|_{\Phi=0} \, .
\ee
From Eq.~\eqref{gaex} we have
\bea \label{gaexf}
&& \Ga[\Phi,\ph_{in}+f]  = \non \\[2mm] \non
 &&  \sum^{\infty}_{n,m=0} \frac{1}{n! m!} \int \!\! \dr^4 x_1 \ldots \int \!\! \dr^4 x_m \int \!\! \dr^4 y_1 \ldots \int \!\! \dr^4 y_n \\[2mm]
&& \times \ga^{m,n}(x_1,\ldots,x_m;y_1,\ldots,y_n) \, \Phi(x_1) \ldots \Phi(x_m) \, : \lf(\ph_{in}(y_1)+f(y_1)\ri) \ldots \lf(\ph_{in}(y_n)+f(y_n)\ri): \, .
\eea
Then
\bea
 \lf. \frac{\de \lan \Ga[\Phi,\ph_{in}+f]\ran}{\de \Phi(x)} \ri|_{\Phi=0}  \ = \ \sum^{\infty}_{n=0} \frac{1}{n!} \int \!\! \dr^4 y_1 \ldots \int \!\! \dr^4 y_n  \ga^{1,n}(x; y_1,\ldots,y_n) \, f(y_1) \ldots f(y_n)\, ,
\eea
so that
\be \label{efdynf1}
\lan\ph^f(x)\ran \ = \ \ - \ \sum^{\infty}_{n=0} \frac{1}{n!} \int \!\! \dr^4 y  \int \!\! \dr^4 y_1 \ldots \int \!\! \dr^4 y_n \, \ga^{1,n}(y; y_1,\ldots,y_n) \, f(y_1) \ldots f(y_n) \,  R(x,y)  \, .
\ee
Taking into account that $\Ga^{(1)}(x)=0$, this relation can be also rewritten as
\be \label{efdynf2}
\Phi^f(x) \ = \ \lan\ph^f(x)\ran \ = \  \sum^{\infty}_{n=1} \frac{1}{n!} \int \!\! \dr^4 y  \int \!\! \dr^4 y_1 \ldots \int \!\! \dr^4 y_n \, \ga^{1,n}(y; y_1,\ldots,y_n) \, f(y_1) \ldots f(y_n) \,  R(x,y)  \, .
\ee
By means of the identification \eqref{ga1n} one can easily prove that this is the same result we could have obtained from Eq.~\eqref{schdyn}. Moreover, $\Phi^f(x)$ is a non-trivial solution of the least-action principle \eqref{gaeqgen}
\be \label{kleq}
\lf. \,\frac{\de \Ga[\Phi]}{\de \Phi(x)}\ri|_{\Phi(x)=\Phi^f(x)} \ = \ 0 \, .
\ee
We now show this in various examples. Let us first apply the previous considerations to the real $\la \ph^4$ theory. From the dynamical map \eqref{treedyn}, after the boson transformation
\be \label{treedynf}
\ph^f(x) \ = \   \lf(\ph_{in}(x)+f(x)\ri)  \ + \ \frac{\la \, }{6} \, \int \!\! \dr^4 y \, \De_R(x-y) \, \lf(\ph_{in}(y)+f(y)\ri)^3 \, . 
\ee
Taking the vacuum expectation value of such expression, one gets
\be
\Phi^f(x)  \ = \     f(x) +\frac{\la \, }{6} \, \int \!\! \dr^4 y \, \De_R(x-y) \, f^3(y) \, .
\ee
This is a solution of \eqref{kleq} at tree level, i.e.
\be
K_x \Phi^f(x) \ = \ \frac{\la \, }{6} \,  \, f^3(x) \ \approx \ \frac{\la }{6} \,  \,\lf(\Phi^f(x)\ri)^3 \, .
\ee

Then, in the case of Eq.~\eqref{treedyn1} one finds
\be
\Phi^f(x) \ = \    f(x)  +\frac{\la \, }{2} \, \int \!\! \dr^4 y \, \De_R(x-y) \, f^2(y) \, .
\ee
Also in this example, Eq.~\eqref{kleq} is satisfied at the tree level
\be
K_x \Phi^f(x) \ = \ \frac{\la \, }{2} \,  \, f^2(x) \ \approx \  \frac{\la }{2} \,  \,\lf(\Phi^f(x)\ri)^2 \, .
\ee
%
%%%%%%%%%%%%%%%%%%%%%%%%%%%%%%%%%%%%%%%%%%%%%%%%%%%
\subsection{Spontaneous symmetry breaking and topological defects}
An important remark is that the condition~\eqref{gmunu} can only be satisfied by massless fields~\cite{Matsumoto:1975rp}. Thus, the natural framework to describe the formation of topological defects are theories with SSB, where Nambu-Goldstone fields can form an inhomogeneous vacuum condensate.

Let us start from the real $\la \ph^4$ theory. In the case $v \neq 0$, the dynamical map is given by Eq.\eqref{treedynssbfinal} instead of Eq.\eqref{treedyn}. Then
\be \label{treedynssbfinalf}
\ph^f(x) \ = \ v +   (\ph_{in}(x)+f(x))   \ + \ \frac{\la \,  \, v}{2} \, \int \!\! \dr^4 y \, \De_R(x-y) \, :(\ph_{in}(y)+f(y))^2: \ + \ \frac{\la \, }{6} \, \int \!\! \dr^4 y \, \De_R(x-y) \, :(\ph_{in}(x)+f(x))^3: \, .
\ee
Therefore
\be \label{treedynssbfinalf1}
\Phi^f(x) \ = \ v +   f(x)   \ + \ \frac{\la \,  \, v}{2} \, \int \!\! \dr^4 y \, \De_R(x-y) \, f^2(y) \ + \ \frac{\la \, }{6} \, \int \!\! \dr^4 y \, \De_R(x-y) \, f^3(x) \, .
\ee
However, in the previous example, the broken symmetry is not a continuous one ($\ph \to -\ph$) and the Goldstone theorem does not apply. In order to study the formation of topological defects in a concrete case, we analyze the complex Goldstone model. If we perform the boson transformation 
\bea \label{bosf}
\chi_{in}(x) & \to & \chi_{in}(x)+f(x) \, ,  \qquad \Box f(x)=0 \, , \\[2mm]
\psi_{in}(x) & \to & \psi_{in}(x)+g(x) \, ,  \qquad (\Box-2\mu^2) g(x)=0 \, , \label{bosg}
\eea 
the dynamical maps \eqref{efdyncomp3},\eqref{efdyncomp4} give
\bea
\chi^f(x)  & = &  (\chi_{in}+f(x))+\frac{\la}{6} \int \!\! \dr^4 y \, \De^\chi_R(x-y) \, :( (\chi_{in}(y)+f(y))^3 \non \\[2mm]
&+& 2 v   (\chi_{in}(y)+f(y)) (\psi_{in}(y)+g(y))+    (\chi_{in}(y)+f(y)) (\psi_{in}(y)+g(y))^2): \, , \\[2mm]
\psi^f(x)  & = &  (\psi_{in}(x)+g(x))+\frac{\la}{6} \int \!\! \dr^4 y \, \De^\psi_R(x-y) \, :\lf[v(3 (\psi_{in}(y)+g(y))^2+  (\chi_{in}(y)+f(y))^2) \non \ri. \\[2mm]
& + & \lf.  (\psi_{in}(y)+g(y)) (\chi_{in}(y)+f(y))^2+(\psi_{in}(y)+g(y))^3\ri]: \, .
\eea
Then
\bea
X^{f,g}(x)  & = &  f(x)+\frac{\la}{6} \int \!\! \dr^4 y \, \De^\chi_R(x-y) \, (f^3(y) + 2 v   f(y) g(y)+   f(y)g^2(y)) \, , \\[2mm]
\Psi^{f,g}(x)  & = &  g(x)+\frac{\la}{6} \int \!\! \dr^4 y \, \De^\psi_R(x-y) \, \lf[v(3 g^2(y)+  f^2(y))  +     g(y) f^2(y)+ g^3(y)\ri] \, .
\eea
In the approximation adopted, these satisfy the equations
\bea
\Box X^{f,g}(x)  & \approx & \frac{\la}{6} ((X^{f,g}(y))^3+2 v X^{f,g}(y) \Psi^{f,g}(y)+ X^{f,g}(y) \Psi^{f,g}(y)) \, , \\[2mm]
(\Box -2 \mu^2)\Psi^{f,g}(x)  & \approx & \frac{\la}{6} \lf[v(3 (\Psi^{f,g}(y))^2+ (X^{f,g}(y))^2)+ \Psi^{f,g}(y) (X^{f,g}(y))^2+(\Psi^{f,g}(y))^3\ri] \, .
\eea
which are equivalent, at tree level, to 
\be
\lf. \frac{\de \Ga[\Psi,X]}{\de \Psi} \ri|_{\Psi=\Psi^{f,g}, X=X^{f,g}} \ = \ 0 \, , \qquad \lf. \frac{\de \Ga[\Psi,X]}{\de X} \ri|_{\Psi=\Psi^{f,g}, X=X^{f,g}} \ = \ 0 \, .
\ee

A topological defect is described by solutions where the condition \eqref{gmunu} is satisfied for $f$, i.e. the Nambu-Goldstone bosons non-trivially condense in the vacuum. In order to better understand this point, let us write the conserved Noether current related to the $U(1)$ symmetry of the Lagrangian
\be \label{noeth}
J_\mu(x) \ = \  \ph^\dag(x) \, \overleftrightarrow{\pa}_\mu \, \ph(x) \, , \qquad \pa^\mu J_\mu \ = \ 0 \, .
\ee
The dynamical maps \eqref{chidyn},\eqref{psidyn} give
\be
\ph(x) \ = \ \frac{v+ \psi_{in}(x)+i  \chi_{in}(x)}{\sqrt{2}} \ + \ \ldots 
\ee
where the dots stand for higher order normal ordered products of $\psi_{in}$ and $\chi_{in}$. If we perform the boson transformation \eqref{bosf} (we now take $g=0$), the vacuum expectation value of the current \eqref{noeth} is
\be
W_\mu(x) \ \equiv \ \lan J_\mu(x)  \ran \ = \ v  \pa_\mu f(x) \, .
\ee
This is a \emph{classical} vector field, whose field strength is non-trivial when the condition $\eqref{gmunu}$ is fulfilled
\be \label{fmunu}
F_{\mu \nu}(x) \ = \ v  \,  \varepsilon_{\mu \nu \rho \si}j^{\rho \si}(x)  \ = \ \pa_\mu W_\nu(x) - \pa_\nu W_\mu(x) \neq 0 \, .
\ee
It is clear that in the above approximation, $f$ can be viewed as a \emph{classical Nambu-Goldstone} field. Similarly as in Refs. \cite{PhysRevD.31.3052,Sakellariadou1990,Battye:1993jv,vilenkin1994cosmic} one can then introduce the $B$-field
\be
W_\mu \ = \ \ha \varepsilon_{\mu \nu \si \rho} \, \pa^\nu B^{\si \rho}(x) 
\ee
Then, the condition Eq.\eqref{gmunu} (or, equivalently \eqref{fmunu}) becomes a wave-equation
\be \label{bwave}
\Box \, B^{\mu \nu}(x) \ = \ j^{\mu \nu}(x) \, .
\ee
For example, with $j^{\mu \nu}$ of the form \eqref{jmunu}, this equation tells us that a string could be the source for a \emph{classical} Nambu-Goldstone radiation described by $B^{\mu \nu}$. This result was previously obtained in the classical approach to global string radiation~\cite{PhysRevD.31.3052, Sakellariadou1990, Battye:1993jv, vilenkin1994cosmic}. However, the present quantum approach is more general and could be used as a framework for a theory of quantum string radiation.  This is a non-trivial task and it will be accomplished in a separate work. Moreover, in the present approach, Eq.\eqref{bwave} could be used to discuss radiation from two or three-dimensional topological objects.

To give a concrete example, one can take $f(x)=\theta=\arctan(x_2/x_1)$, which solves $\Box \theta =\nabla^2 \theta=0$. Then  $\chi^f$ and $\psi^f$ are solutions of the field equations when a linear static-string solution along the $x_3$-axis is formed \cite{Acquaviva:2020cjx}. In fact, in that case
\be
j^{0 3}(x) \ = \ -j^{3 0}(x) \ = \ \de(x_1) \de(x_2) \, ,
\ee
which means $y^0=\tau$, $y^1=y^2=0$ and $y^3 = \si$ (see Eq.\eqref{jmunu}).
%%%%%%%%%%%%%%%%%%%%%%%%%%%%%%%%%%%%%%%%%%%%%%%%%%%%%%%%%%%%%%%%%%%%%%%%%
\section{Conclusions} \label{sec5}
In this work, we built an effective action formulation of the dynamical map of a scalar field. The result was achieved by defining an operator-valued effective action, whose vacuum expectation value is the generating functional of retarded, one-particle irreducible, $n$-point functions. The formalism is general and can be applied to systems underlying SSB. Moreover, the usual loop expansion of the effective action can be employed to compute higher-order results. Then we showed how a boson transformation (inhomogeneous condensation in vacuum) acts on such a mapping and thus we created a link between Umezawa's boson method and the common approach to deal with topological defects, based on the least-action principle.

Our work is a first step towards a description allowing a fully quantum treatment of radiation from topological defects. Although the present paper describes the case of a scalar field, it can be generalized to other situations. In particular to gauge theories and the cases where bound states appear in the physical spectrum. This would allow a quantum field theory treatment of cosmic string radiation, which is usually studied within a semi-classical framework.

%Although the present paper describes the case of a scalar field we plan to generalize it in the future to other situations. In particular, to gauge theories and the cases where bound states appear in the physical spectrum. These generalizations would permit the treatment of more realistic cases where topological defects are usually encountered. %For example, as mentioned in the main text, the present work could be a starting point for a quantum treatment of cosmic string radiation, which is usually studied within a classical framework, in analogy with the theory of radiation from antennas in classical electrodynamics \cite{PhysRevD.31.3052,Sakellariadou1990,Battye:1993jv,vilenkin1994cosmic}.
%%%%%%%%%%%%%%%%%%%%%%%%%%%%%%%%%%%%%%%%%%%%%%%%%%%%%%%%%%%%%%%%%%%%%%%%
\section*{Acknowledgements}
%This work was supported by the Polish National Science Center grant 2018/31/D/ST2/02048.
This work was supported by the Polish National Agency for Academic Exchange
within Polish Returns Programme under agreement PPN/PPO/2020/1/00013/U/00001 and
the Polish National Science Center grant 2018/31/D/ST2/02048. L.S. would like to thank the useful comments of M. Blasone and G. Vitiello.
%%%%%%%%%%%%%%%%%%%%%%%%%%%%%%%%%%%%%%%%%%%%%%%%%%%%%%%%%%%%%%%%%%%%%%%%
%%%%%%%%%%%%%%%%%%%%%%%%%%%%%%%%%%%%%%%%%%%%%%%%%%%%%%%%%%%%%%%%%%%%%%%%
%\section*{References}

\bibliographystyle{apsrev4-2}
\bibliography{libraryCosmo}

\begin{thebibliography}{48}%
\makeatletter
\providecommand \@ifxundefined [1]{%
 \@ifx{#1\undefined}
}%
\providecommand \@ifnum [1]{%
 \ifnum #1\expandafter \@firstoftwo
 \else \expandafter \@secondoftwo
 \fi
}%
\providecommand \@ifx [1]{%
 \ifx #1\expandafter \@firstoftwo
 \else \expandafter \@secondoftwo
 \fi
}%
\providecommand \natexlab [1]{#1}%
\providecommand \enquote  [1]{``#1''}%
\providecommand \bibnamefont  [1]{#1}%
\providecommand \bibfnamefont [1]{#1}%
\providecommand \citenamefont [1]{#1}%
\providecommand \href@noop [0]{\@secondoftwo}%
\providecommand \href [0]{\begingroup \@sanitize@url \@href}%
\providecommand \@href[1]{\@@startlink{#1}\@@href}%
\providecommand \@@href[1]{\endgroup#1\@@endlink}%
\providecommand \@sanitize@url [0]{\catcode `\\12\catcode `\$12\catcode
  `\&12\catcode `\#12\catcode `\^12\catcode `\_12\catcode `\%12\relax}%
\providecommand \@@startlink[1]{}%
\providecommand \@@endlink[0]{}%
\providecommand \url  [0]{\begingroup\@sanitize@url \@url }%
\providecommand \@url [1]{\endgroup\@href {#1}{\urlprefix }}%
\providecommand \urlprefix  [0]{URL }%
\providecommand \Eprint [0]{\href }%
\providecommand \doibase [0]{https://doi.org/}%
\providecommand \selectlanguage [0]{\@gobble}%
\providecommand \bibinfo  [0]{\@secondoftwo}%
\providecommand \bibfield  [0]{\@secondoftwo}%
\providecommand \translation [1]{[#1]}%
\providecommand \BibitemOpen [0]{}%
\providecommand \bibitemStop [0]{}%
\providecommand \bibitemNoStop [0]{.\EOS\space}%
\providecommand \EOS [0]{\spacefactor3000\relax}%
\providecommand \BibitemShut  [1]{\csname bibitem#1\endcsname}%
\let\auto@bib@innerbib\@empty
%</preamble>
\bibitem [{\citenamefont {Itzykson}\ and\ \citenamefont
  {Zuber}(2012)}]{itzykson2012quantum}%
  \BibitemOpen
  \bibfield  {author} {\bibinfo {author} {\bibfnamefont {C.}~\bibnamefont
  {Itzykson}}\ and\ \bibinfo {author} {\bibfnamefont {J.}~\bibnamefont
  {Zuber}},\ }\href {https://books.google.it/books?id=CxYCMNrUnTEC} {\emph
  {\bibinfo {title} {Quantum Field Theory}}},\ Dover Books on Physics\
  (\bibinfo  {publisher} {Dover Publications},\ \bibinfo {year}
  {2012})\BibitemShut {NoStop}%
\bibitem [{\citenamefont {Kleinert}(2016)}]{kleinert2016particles}%
  \BibitemOpen
  \bibfield  {author} {\bibinfo {author} {\bibfnamefont {H.}~\bibnamefont
  {Kleinert}},\ }\href {https://books.google.it/books?id=d1-2DAAAQBAJ} {\emph
  {\bibinfo {title} {Particles And Quantum Fields}}}\ (\bibinfo  {publisher}
  {World Scientific Publishing Company},\ \bibinfo {year} {2016})\BibitemShut
  {NoStop}%
\bibitem [{\citenamefont {Schweber}(2013)}]{schweber2013introduction}%
  \BibitemOpen
  \bibfield  {author} {\bibinfo {author} {\bibfnamefont {S.}~\bibnamefont
  {Schweber}},\ }\href {https://books.google.it/books?id=v1owGsfiJcoC} {\emph
  {\bibinfo {title} {An Introduction to Relativistic Quantum Field Theory}}}\
  (\bibinfo  {publisher} {Dover Publications},\ \bibinfo {year}
  {2013})\BibitemShut {NoStop}%
\bibitem [{\citenamefont {Haag}(1955)}]{Haag:1955ev}%
  \BibitemOpen
  \bibfield  {author} {\bibinfo {author} {\bibfnamefont {R.}~\bibnamefont
  {Haag}},\ }\href@noop {} {\bibfield  {journal} {\bibinfo  {journal} {Kong.
  Dan. Vid. Sel. Mat. Fys. Med.}\ }\textbf {\bibinfo {volume} {29N12}},\
  \bibinfo {pages} {1} (\bibinfo {year} {1955})}\BibitemShut {NoStop}%
\bibitem [{\citenamefont {Umezawa}\ \emph {et~al.}(1982)\citenamefont
  {Umezawa}, \citenamefont {Matsumoto},\ and\ \citenamefont
  {Tachiki}}]{Umezawa:1982nv}%
  \BibitemOpen
  \bibfield  {author} {\bibinfo {author} {\bibfnamefont {H.}~\bibnamefont
  {Umezawa}}, \bibinfo {author} {\bibfnamefont {H.}~\bibnamefont {Matsumoto}},\
  and\ \bibinfo {author} {\bibfnamefont {M.}~\bibnamefont {Tachiki}},\
  }\href@noop {} {\emph {\bibinfo {title} {{Thermo-Field Dynamics and Condensed
  States}}}}\ (\bibinfo  {publisher} {North-Holland},\ \bibinfo {year}
  {1982})\BibitemShut {NoStop}%
%%CITATION = INSPIRE-185489;%%
\bibitem [{\citenamefont {Umezawa}(1993)}]{Umezawa:1993yq}%
  \BibitemOpen
  \bibfield  {author} {\bibinfo {author} {\bibfnamefont {H.}~\bibnamefont
  {Umezawa}},\ }\href@noop {} {\emph {\bibinfo {title} {{Advanced field theory:
  Micro, macro, and thermal physics}}}}\ (\bibinfo  {publisher} {AIP},\
  \bibinfo {year} {1993})\BibitemShut {NoStop}%
%%CITATION = INSPIRE-362835;%%
\bibitem [{\citenamefont {Greenberg}(1994)}]{Greenberg:1994zu}%
  \BibitemOpen
  \bibfield  {author} {\bibinfo {author} {\bibfnamefont {O.~W.}\ \bibnamefont
  {Greenberg}},\ }in\ \href@noop {} {\emph {\bibinfo {booktitle} {{1st Arctic
  Workshop on Future Physics and Accelerators}}}}\ (\bibinfo {year} {1994})\
  pp.\ \bibinfo {pages} {0498--520},\ \Eprint
  {https://arxiv.org/abs/hep-ph/9502253} {arXiv:hep-ph/9502253} \BibitemShut
  {NoStop}%
\bibitem [{\citenamefont {{Blasone}}\ \emph {et~al.}(2011)\citenamefont
  {{Blasone}}, \citenamefont {{Jizba}},\ and\ \citenamefont
  {{Vitiello}}}]{Blasone2011}%
  \BibitemOpen
  \bibfield  {author} {\bibinfo {author} {\bibfnamefont {M.}~\bibnamefont
  {{Blasone}}}, \bibinfo {author} {\bibfnamefont {P.}~\bibnamefont {{Jizba}}},\
  and\ \bibinfo {author} {\bibfnamefont {G.}~\bibnamefont {{Vitiello}}},\
  }\href {https://books.google.cz/books?id=diVZqQAVg40C} {\emph {\bibinfo
  {title} {{Quantum Field Theory and Its Macroscopic Manifestations: Boson
  Condensation, Ordered Patterns, and Topological Defects}}}}\ (\bibinfo
  {publisher} {Imperial College Press},\ \bibinfo {year} {2011})\BibitemShut
  {NoStop}%
\bibitem [{\citenamefont {Umezawa}(1965)}]{Umezawa1965DynamicalRO}%
  \BibitemOpen
  \bibfield  {author} {\bibinfo {author} {\bibfnamefont {H.}~\bibnamefont
  {Umezawa}},\ }\href@noop {} {\bibfield  {journal} {\bibinfo  {journal} {Il
  Nuovo Cimento A (1965-1970)}\ }\textbf {\bibinfo {volume} {40}},\ \bibinfo
  {pages} {450} (\bibinfo {year} {1965})}\BibitemShut {NoStop}%
\bibitem [{\citenamefont {Matsumoto}\ \emph {et~al.}(1974)\citenamefont
  {Matsumoto}, \citenamefont {Papastamatiou},\ and\ \citenamefont
  {Umezawa}}]{Matsumoto:1973hg}%
  \BibitemOpen
  \bibfield  {author} {\bibinfo {author} {\bibfnamefont {H.}~\bibnamefont
  {Matsumoto}}, \bibinfo {author} {\bibfnamefont {N.~J.}\ \bibnamefont
  {Papastamatiou}},\ and\ \bibinfo {author} {\bibfnamefont {H.}~\bibnamefont
  {Umezawa}},\ }\href {https://doi.org/10.1016/0550-3213(74)90578-1} {\bibfield
   {journal} {\bibinfo  {journal} {Nucl. Phys. B}\ }\textbf {\bibinfo {volume}
  {{ 82}}},\ \bibinfo {pages} {45} (\bibinfo {year} {1974})}\BibitemShut
  {NoStop}%
%%CITATION = NUPHA,B82,45;%%
\bibitem [{\citenamefont {Matsumoto}\ \emph
  {et~al.}(1975{\natexlab{a}})\citenamefont {Matsumoto}, \citenamefont
  {Papastamatiou}, \citenamefont {Umezawa},\ and\ \citenamefont
  {Vitiello}}]{Matsumoto:1975fi}%
  \BibitemOpen
  \bibfield  {author} {\bibinfo {author} {\bibfnamefont {H.}~\bibnamefont
  {Matsumoto}}, \bibinfo {author} {\bibfnamefont {N.}~\bibnamefont
  {Papastamatiou}}, \bibinfo {author} {\bibfnamefont {H.}~\bibnamefont
  {Umezawa}},\ and\ \bibinfo {author} {\bibfnamefont {G.}~\bibnamefont
  {Vitiello}},\ }\href {https://doi.org/10.1016/0550-3213(75)90215-1}
  {\bibfield  {journal} {\bibinfo  {journal} {Nucl. Phys. B}\ }\textbf
  {\bibinfo {volume} {{ 97}}},\ \bibinfo {pages} {61} (\bibinfo {year}
  {1975}{\natexlab{a}})}\BibitemShut {NoStop}%
\bibitem [{\citenamefont {De~Concini}\ and\ \citenamefont
  {Vitiello}(1976)}]{DeConcini:1976uk}%
  \BibitemOpen
  \bibfield  {author} {\bibinfo {author} {\bibfnamefont {C.}~\bibnamefont
  {De~Concini}}\ and\ \bibinfo {author} {\bibfnamefont {G.}~\bibnamefont
  {Vitiello}},\ }\href {https://doi.org/10.1016/0550-3213(76)90317-5}
  {\bibfield  {journal} {\bibinfo  {journal} {Nucl. Phys. B}\ }\textbf
  {\bibinfo {volume} {116}},\ \bibinfo {pages} {141} (\bibinfo {year}
  {1976})}\BibitemShut {NoStop}%
\bibitem [{\citenamefont {Hongoh}\ \emph {et~al.}(1981)\citenamefont {Hongoh},
  \citenamefont {Matsumoto},\ and\ \citenamefont
  {Umezawa}}]{10.1143/PTP.65.315}%
  \BibitemOpen
  \bibfield  {author} {\bibinfo {author} {\bibfnamefont {M.}~\bibnamefont
  {Hongoh}}, \bibinfo {author} {\bibfnamefont {H.}~\bibnamefont {Matsumoto}},\
  and\ \bibinfo {author} {\bibfnamefont {H.}~\bibnamefont {Umezawa}},\ }\href
  {https://doi.org/10.1143/PTP.65.315} {\bibfield  {journal} {\bibinfo
  {journal} {Progress of Theoretical Physics}\ }\textbf {\bibinfo {volume}
  {65}},\ \bibinfo {pages} {315} (\bibinfo {year} {1981})}\BibitemShut
  {NoStop}%
\bibitem [{\citenamefont {Wadati}\ \emph {et~al.}(1977)\citenamefont {Wadati},
  \citenamefont {Matsumoto}, \citenamefont {Takahashi},\ and\ \citenamefont
  {Umezawa}}]{Wadati1977ASF}%
  \BibitemOpen
  \bibfield  {author} {\bibinfo {author} {\bibfnamefont {M.}~\bibnamefont
  {Wadati}}, \bibinfo {author} {\bibfnamefont {H.}~\bibnamefont {Matsumoto}},
  \bibinfo {author} {\bibfnamefont {Y.}~\bibnamefont {Takahashi}},\ and\
  \bibinfo {author} {\bibfnamefont {H.}~\bibnamefont {Umezawa}},\ }\href@noop
  {} {\bibfield  {journal} {\bibinfo  {journal} {Physics Letters A}\ }\textbf
  {\bibinfo {volume} {62}},\ \bibinfo {pages} {255} (\bibinfo {year}
  {1977})}\BibitemShut {NoStop}%
\bibitem [{\citenamefont {Wadati}\ \emph
  {et~al.}(1978{\natexlab{a}})\citenamefont {Wadati}, \citenamefont
  {Matsumoto}, \citenamefont {Takahashi},\ and\ \citenamefont
  {Umezawa}}]{Wadati1978FDP}%
  \BibitemOpen
  \bibfield  {author} {\bibinfo {author} {\bibfnamefont {M.}~\bibnamefont
  {Wadati}}, \bibinfo {author} {\bibfnamefont {H.}~\bibnamefont {Matsumoto}},
  \bibinfo {author} {\bibfnamefont {Y.}~\bibnamefont {Takahashi}},\ and\
  \bibinfo {author} {\bibfnamefont {H.}~\bibnamefont {Umezawa}},\ }\href
  {https://doi.org/https://doi.org/10.1002/prop.19780260602} {\bibfield
  {journal} {\bibinfo  {journal} {Fortschr. Phys.}\ }\textbf {\bibinfo {volume}
  {26}},\ \bibinfo {pages} {357} (\bibinfo {year}
  {1978}{\natexlab{a}})}\BibitemShut {NoStop}%
\bibitem [{\citenamefont {Wadati}\ \emph
  {et~al.}(1978{\natexlab{b}})\citenamefont {Wadati}, \citenamefont
  {Matsumoto},\ and\ \citenamefont {Umezawa}}]{PhysRevB.18.4077}%
  \BibitemOpen
  \bibfield  {author} {\bibinfo {author} {\bibfnamefont {M.}~\bibnamefont
  {Wadati}}, \bibinfo {author} {\bibfnamefont {H.}~\bibnamefont {Matsumoto}},\
  and\ \bibinfo {author} {\bibfnamefont {H.}~\bibnamefont {Umezawa}},\ }\href
  {https://doi.org/10.1103/PhysRevB.18.4077} {\bibfield  {journal} {\bibinfo
  {journal} {Phys. Rev. B}\ }\textbf {\bibinfo {volume} {18}},\ \bibinfo
  {pages} {4077} (\bibinfo {year} {1978}{\natexlab{b}})}\BibitemShut {NoStop}%
\bibitem [{\citenamefont {Mercaldo}\ \emph {et~al.}(1981)\citenamefont
  {Mercaldo}, \citenamefont {Rabuffo},\ and\ \citenamefont
  {Vitiello}}]{MERCALDO1981193}%
  \BibitemOpen
  \bibfield  {author} {\bibinfo {author} {\bibfnamefont {L.}~\bibnamefont
  {Mercaldo}}, \bibinfo {author} {\bibfnamefont {I.}~\bibnamefont {Rabuffo}},\
  and\ \bibinfo {author} {\bibfnamefont {G.}~\bibnamefont {Vitiello}},\ }\href
  {https://doi.org/https://doi.org/10.1016/0550-3213(81)90112-7} {\bibfield
  {journal} {\bibinfo  {journal} {Nuclear Physics B}\ }\textbf {\bibinfo
  {volume} {188}},\ \bibinfo {pages} {193} (\bibinfo {year}
  {1981})}\BibitemShut {NoStop}%
\bibitem [{\citenamefont {Blasone}\ and\ \citenamefont
  {Jizba}(2002)}]{Blasone:2001aj}%
  \BibitemOpen
  \bibfield  {author} {\bibinfo {author} {\bibfnamefont {M.}~\bibnamefont
  {Blasone}}\ and\ \bibinfo {author} {\bibfnamefont {P.}~\bibnamefont
  {Jizba}},\ }\href {https://doi.org/10.1006/aphy.2001.6215} {\bibfield
  {journal} {\bibinfo  {journal} {Ann. Phys.}\ }\textbf {\bibinfo {volume} {{
  295}}},\ \bibinfo {pages} {230} (\bibinfo {year} {2002})},\ \Eprint
  {https://arxiv.org/abs/hep-th/0108177} {arXiv:hep-th/0108177} \BibitemShut
  {NoStop}%
\bibitem [{\citenamefont {Leplae}\ and\ \citenamefont
  {Umezawa}(1969)}]{LepUme}%
  \BibitemOpen
  \bibfield  {author} {\bibinfo {author} {\bibfnamefont {L.}~\bibnamefont
  {Leplae}}\ and\ \bibinfo {author} {\bibfnamefont {H.}~\bibnamefont
  {Umezawa}},\ }\href@noop {} {\bibfield  {journal} {\bibinfo  {journal} {J.
  Math. Phys.}\ }\textbf {\bibinfo {volume} {{ 10}}},\ \bibinfo {pages} {2038}
  (\bibinfo {year} {1969})}\BibitemShut {NoStop}%
\bibitem [{\citenamefont {Leplae}\ \emph {et~al.}(1970)\citenamefont {Leplae},
  , \citenamefont {Mancini},\ and\ \citenamefont {Umezawa}}]{LepManUme}%
  \BibitemOpen
  \bibfield  {author} {\bibinfo {author} {\bibfnamefont {L.}~\bibnamefont
  {Leplae}}, , \bibinfo {author} {\bibfnamefont {F.}~\bibnamefont {Mancini}},\
  and\ \bibinfo {author} {\bibfnamefont {H.}~\bibnamefont {Umezawa}},\
  }\href@noop {} {\bibfield  {journal} {\bibinfo  {journal} {Phys. Rev. B}\
  }\textbf {\bibinfo {volume} {{ 2}}},\ \bibinfo {pages} {3594} (\bibinfo
  {year} {1970})}\BibitemShut {NoStop}%
\bibitem [{\citenamefont {Tze}\ and\ \citenamefont {Ezawa}(1975)}]{TZE197563}%
  \BibitemOpen
  \bibfield  {author} {\bibinfo {author} {\bibfnamefont {H.}~\bibnamefont
  {Tze}}\ and\ \bibinfo {author} {\bibfnamefont {Z.}~\bibnamefont {Ezawa}},\
  }\href {https://doi.org/https://doi.org/10.1016/0370-2693(75)90188-4}
  {\bibfield  {journal} {\bibinfo  {journal} {Physics Letters B}\ }\textbf
  {\bibinfo {volume} {{ 55}}},\ \bibinfo {pages} {63} (\bibinfo {year}
  {1975})}\BibitemShut {NoStop}%
\bibitem [{\citenamefont {Wadati}\ \emph
  {et~al.}(1978{\natexlab{c}})\citenamefont {Wadati}, \citenamefont
  {Matsumoto},\ and\ \citenamefont {Umezawa}}]{PhysRevD.18.1192}%
  \BibitemOpen
  \bibfield  {author} {\bibinfo {author} {\bibfnamefont {M.}~\bibnamefont
  {Wadati}}, \bibinfo {author} {\bibfnamefont {H.}~\bibnamefont {Matsumoto}},\
  and\ \bibinfo {author} {\bibfnamefont {H.}~\bibnamefont {Umezawa}},\ }\href
  {https://doi.org/10.1103/PhysRevD.18.1192} {\bibfield  {journal} {\bibinfo
  {journal} {Phys. Rev. D}\ }\textbf {\bibinfo {volume} {18}},\ \bibinfo
  {pages} {1192} (\bibinfo {year} {1978}{\natexlab{c}})}\BibitemShut {NoStop}%
\bibitem [{\citenamefont {Iorio}\ and\ \citenamefont
  {Smaldone}(2023{\natexlab{a}})}]{Iorio2023}%
  \BibitemOpen
  \bibfield  {author} {\bibinfo {author} {\bibfnamefont {A.}~\bibnamefont
  {Iorio}}\ and\ \bibinfo {author} {\bibfnamefont {L.}~\bibnamefont
  {Smaldone}},\ }\href {https://doi.org/10.1142/S0218271823500633} {\bibfield
  {journal} {\bibinfo  {journal} {Int. J. Mod. Phys. D}\ }\textbf {\bibinfo
  {volume} {32}},\ \bibinfo {pages} {2350063} (\bibinfo {year}
  {2023}{\natexlab{a}})}\BibitemShut {NoStop}%
\bibitem [{\citenamefont {Iorio}\ and\ \citenamefont
  {Smaldone}(2023{\natexlab{b}})}]{Iorio:2023sav}%
  \BibitemOpen
  \bibfield  {author} {\bibinfo {author} {\bibfnamefont {A.}~\bibnamefont
  {Iorio}}\ and\ \bibinfo {author} {\bibfnamefont {L.}~\bibnamefont
  {Smaldone}},\ }\href {https://doi.org/10.1088/1742-6596/2533/1/012030}
  {\bibfield  {journal} {\bibinfo  {journal} {J. Phys. Conf. Ser.}\ }\textbf
  {\bibinfo {volume} {2533}},\ \bibinfo {pages} {012030} (\bibinfo {year}
  {2023}{\natexlab{b}})}\BibitemShut {NoStop}%
\bibitem [{\citenamefont {Mańka}\ \emph {et~al.}(1986)\citenamefont {Mańka},
  \citenamefont {Kuczyński},\ and\ \citenamefont {Vitiello}}]{MANKA1986533}%
  \BibitemOpen
  \bibfield  {author} {\bibinfo {author} {\bibfnamefont {R.}~\bibnamefont
  {Mańka}}, \bibinfo {author} {\bibfnamefont {J.}~\bibnamefont {Kuczyński}},\
  and\ \bibinfo {author} {\bibfnamefont {G.}~\bibnamefont {Vitiello}},\ }\href
  {https://doi.org/https://doi.org/10.1016/0550-3213(86)90064-7} {\bibfield
  {journal} {\bibinfo  {journal} {Nuclear Physics B}\ }\textbf {\bibinfo
  {volume} {276}},\ \bibinfo {pages} {533} (\bibinfo {year}
  {1986})}\BibitemShut {NoStop}%
\bibitem [{\citenamefont {Mańka}\ and\ \citenamefont
  {Vitiello}(1990)}]{MANKA199061}%
  \BibitemOpen
  \bibfield  {author} {\bibinfo {author} {\bibfnamefont {R.}~\bibnamefont
  {Mańka}}\ and\ \bibinfo {author} {\bibfnamefont {G.}~\bibnamefont
  {Vitiello}},\ }\href
  {https://doi.org/https://doi.org/10.1016/0003-4916(90)90368-X} {\bibfield
  {journal} {\bibinfo  {journal} {Annals of Physics}\ }\textbf {\bibinfo
  {volume} {199}},\ \bibinfo {pages} {61} (\bibinfo {year} {1990})}\BibitemShut
  {NoStop}%
\bibitem [{\citenamefont {Swanson}(1981)}]{Swanson1981}%
  \BibitemOpen
  \bibfield  {author} {\bibinfo {author} {\bibfnamefont {M.~S.}\ \bibnamefont
  {Swanson}},\ }\href {https://doi.org/https://doi.org/10.1063/1.524983}
  {\bibfield  {journal} {\bibinfo  {journal} {J. Math. Phys.}\ }\textbf
  {\bibinfo {volume} {22}},\ \bibinfo {pages} {777} (\bibinfo {year}
  {1981})}\BibitemShut {NoStop}%
\bibitem [{\citenamefont {Yang}\ and\ \citenamefont
  {Feldman}(1950)}]{Yang:1950vi}%
  \BibitemOpen
  \bibfield  {author} {\bibinfo {author} {\bibfnamefont {C.-N.}\ \bibnamefont
  {Yang}}\ and\ \bibinfo {author} {\bibfnamefont {D.}~\bibnamefont {Feldman}},\
  }\href {https://doi.org/10.1103/PhysRev.79.972} {\bibfield  {journal}
  {\bibinfo  {journal} {Phys. Rev.}\ }\textbf {\bibinfo {volume} {{ 79}}},\
  \bibinfo {pages} {972} (\bibinfo {year} {1950})}\BibitemShut {NoStop}%
\bibitem [{\citenamefont {Greiner}\ \emph {et~al.}(2013)\citenamefont
  {Greiner}, \citenamefont {Bromley},\ and\ \citenamefont
  {Reinhardt}}]{greiner2013field}%
  \BibitemOpen
  \bibfield  {author} {\bibinfo {author} {\bibfnamefont {W.}~\bibnamefont
  {Greiner}}, \bibinfo {author} {\bibfnamefont {D.}~\bibnamefont {Bromley}},\
  and\ \bibinfo {author} {\bibfnamefont {J.}~\bibnamefont {Reinhardt}},\ }\href
  {https://books.google.pl/books?id=C-DVBAAAQBAJ} {\emph {\bibinfo {title}
  {Field Quantization}}}\ (\bibinfo  {publisher} {Springer Berlin Heidelberg},\
  \bibinfo {year} {2013})\BibitemShut {NoStop}%
\bibitem [{\citenamefont {Ryder}(1996)}]{ryder1996}%
  \BibitemOpen
  \bibfield  {author} {\bibinfo {author} {\bibfnamefont {L.~H.}\ \bibnamefont
  {Ryder}},\ }\href {https://doi.org/10.1017/CBO9780511813900} {\emph {\bibinfo
  {title} {Quantum Field Theory}}},\ \bibinfo {edition} {2nd}\ ed.\ (\bibinfo
  {publisher} {Cambridge University Press},\ \bibinfo {year}
  {1996})\BibitemShut {NoStop}%
\bibitem [{\citenamefont {Greiner}\ and\ \citenamefont
  {Reinhardt}(2003)}]{greiner2003quantum}%
  \BibitemOpen
  \bibfield  {author} {\bibinfo {author} {\bibfnamefont {W.}~\bibnamefont
  {Greiner}}\ and\ \bibinfo {author} {\bibfnamefont {J.}~\bibnamefont
  {Reinhardt}},\ }\href {https://books.google.it/books?id=Ci-9XMwzkmoC} {\emph
  {\bibinfo {title} {Quantum Electrodynamics}}},\ Physics and astronomy online
  library\ (\bibinfo  {publisher} {Springer},\ \bibinfo {year}
  {2003})\BibitemShut {NoStop}%
\bibitem [{\citenamefont {Jevicki}\ and\ \citenamefont
  {Lee}(1988)}]{Jevicki:1987ax}%
  \BibitemOpen
  \bibfield  {author} {\bibinfo {author} {\bibfnamefont {A.}~\bibnamefont
  {Jevicki}}\ and\ \bibinfo {author} {\bibfnamefont {C.-k.}\ \bibnamefont
  {Lee}},\ }\href {https://doi.org/10.1103/PhysRevD.37.1485} {\bibfield
  {journal} {\bibinfo  {journal} {Phys. Rev. D}\ }\textbf {\bibinfo {volume}
  {37}},\ \bibinfo {pages} {1485} (\bibinfo {year} {1988})}\BibitemShut
  {NoStop}%
\bibitem [{\citenamefont {Brandt}\ \emph {et~al.}(1999)\citenamefont {Brandt},
  \citenamefont {Das}, \citenamefont {Frenkel},\ and\ \citenamefont
  {da~Silva}}]{PhysRevD.59.065004}%
  \BibitemOpen
  \bibfield  {author} {\bibinfo {author} {\bibfnamefont {F.~T.}\ \bibnamefont
  {Brandt}}, \bibinfo {author} {\bibfnamefont {A.}~\bibnamefont {Das}},
  \bibinfo {author} {\bibfnamefont {J.}~\bibnamefont {Frenkel}},\ and\ \bibinfo
  {author} {\bibfnamefont {A.~J.}\ \bibnamefont {da~Silva}},\ }\href
  {https://doi.org/10.1103/PhysRevD.59.065004} {\bibfield  {journal} {\bibinfo
  {journal} {Phys. Rev. D}\ }\textbf {\bibinfo {volume} {59}},\ \bibinfo
  {pages} {065004} (\bibinfo {year} {1999})}\BibitemShut {NoStop}%
\bibitem [{\citenamefont {Goldstone}(1961)}]{Goldstone:1961eq}%
  \BibitemOpen
  \bibfield  {author} {\bibinfo {author} {\bibfnamefont {J.}~\bibnamefont
  {Goldstone}},\ }\href {https://doi.org/10.1007/BF02812722} {\bibfield
  {journal} {\bibinfo  {journal} {Nuovo Cim.}\ }\textbf {\bibinfo {volume}
  {19}},\ \bibinfo {pages} {154} (\bibinfo {year} {1961})}\BibitemShut
  {NoStop}%
\bibitem [{\citenamefont {Goldstone}\ \emph {et~al.}(1962)\citenamefont
  {Goldstone}, \citenamefont {Salam},\ and\ \citenamefont
  {Weinberg}}]{PhysRev.127.965}%
  \BibitemOpen
  \bibfield  {author} {\bibinfo {author} {\bibfnamefont {J.}~\bibnamefont
  {Goldstone}}, \bibinfo {author} {\bibfnamefont {A.}~\bibnamefont {Salam}},\
  and\ \bibinfo {author} {\bibfnamefont {S.}~\bibnamefont {Weinberg}},\ }\href
  {https://doi.org/10.1103/PhysRev.127.965} {\bibfield  {journal} {\bibinfo
  {journal} {Phys. Rev.}\ }\textbf {\bibinfo {volume} {127}},\ \bibinfo {pages}
  {965} (\bibinfo {year} {1962})}\BibitemShut {NoStop}%
\bibitem [{\citenamefont {Peskin}\ and\ \citenamefont
  {Schroeder}(1995)}]{peskin2018introduction}%
  \BibitemOpen
  \bibfield  {author} {\bibinfo {author} {\bibfnamefont {M.}~\bibnamefont
  {Peskin}}\ and\ \bibinfo {author} {\bibfnamefont {D.}~\bibnamefont
  {Schroeder}},\ }\href {https://books.google.it/books?id=EVeNNcslvX0C} {\emph
  {\bibinfo {title} {An Introduction To Quantum Field Theory}}},\ Frontiers in
  Physics\ (\bibinfo  {publisher} {Avalon Publishing},\ \bibinfo {year}
  {1995})\BibitemShut {NoStop}%
\bibitem [{\citenamefont {Cheyette}(1985)}]{Cheyette:1985ue}%
  \BibitemOpen
  \bibfield  {author} {\bibinfo {author} {\bibfnamefont {O.}~\bibnamefont
  {Cheyette}},\ }\href {https://doi.org/10.1103/PhysRevLett.55.2394} {\bibfield
   {journal} {\bibinfo  {journal} {Phys. Rev. Lett.}\ }\textbf {\bibinfo
  {volume} {55}},\ \bibinfo {pages} {2394} (\bibinfo {year}
  {1985})}\BibitemShut {NoStop}%
\bibitem [{\citenamefont {Chan}(1986)}]{PhysRevLett.57.1199}%
  \BibitemOpen
  \bibfield  {author} {\bibinfo {author} {\bibfnamefont {L.-H.}\ \bibnamefont
  {Chan}},\ }\href {https://doi.org/10.1103/PhysRevLett.57.1199} {\bibfield
  {journal} {\bibinfo  {journal} {Phys. Rev. Lett.}\ }\textbf {\bibinfo
  {volume} {57}},\ \bibinfo {pages} {1199} (\bibinfo {year}
  {1986})}\BibitemShut {NoStop}%
\bibitem [{\citenamefont {Buchbinder}\ \emph {et~al.}(1992)\citenamefont
  {Buchbinder}, \citenamefont {Odintsov},\ and\ \citenamefont
  {Shapiro}}]{buchbinder1992effective}%
  \BibitemOpen
  \bibfield  {author} {\bibinfo {author} {\bibfnamefont {I.}~\bibnamefont
  {Buchbinder}}, \bibinfo {author} {\bibfnamefont {S.}~\bibnamefont
  {Odintsov}},\ and\ \bibinfo {author} {\bibfnamefont {L.}~\bibnamefont
  {Shapiro}},\ }\href {https://books.google.it/books?id=NcjI3ydY4e4C} {\emph
  {\bibinfo {title} {Effective Action in Quantum Gravity}}}\ (\bibinfo
  {publisher} {Taylor \& Francis},\ \bibinfo {year} {1992})\BibitemShut
  {NoStop}%
\bibitem [{\citenamefont {Coleman}\ and\ \citenamefont
  {Weinberg}(1973)}]{PhysRevD.7.1888}%
  \BibitemOpen
  \bibfield  {author} {\bibinfo {author} {\bibfnamefont {S.}~\bibnamefont
  {Coleman}}\ and\ \bibinfo {author} {\bibfnamefont {E.}~\bibnamefont
  {Weinberg}},\ }\href {https://doi.org/10.1103/PhysRevD.7.1888} {\bibfield
  {journal} {\bibinfo  {journal} {Phys. Rev. D}\ }\textbf {\bibinfo {volume}
  {7}},\ \bibinfo {pages} {1888} (\bibinfo {year} {1973})}\BibitemShut
  {NoStop}%
\bibitem [{\citenamefont {Matsumoto}\ \emph
  {et~al.}(1975{\natexlab{b}})\citenamefont {Matsumoto}, \citenamefont
  {Papastamatiou},\ and\ \citenamefont {Umezawa}}]{Matsumoto:1975rp}%
  \BibitemOpen
  \bibfield  {author} {\bibinfo {author} {\bibfnamefont {H.}~\bibnamefont
  {Matsumoto}}, \bibinfo {author} {\bibfnamefont {N.}~\bibnamefont
  {Papastamatiou}},\ and\ \bibinfo {author} {\bibfnamefont {H.}~\bibnamefont
  {Umezawa}},\ }\href {https://doi.org/10.1016/0550-3213(75)90216-3} {\bibfield
   {journal} {\bibinfo  {journal} {Nucl. Phys. B}\ }\textbf {\bibinfo {volume}
  {97}},\ \bibinfo {pages} {90} (\bibinfo {year}
  {1975}{\natexlab{b}})}\BibitemShut {NoStop}%
\bibitem [{\citenamefont {Ezawa}\ and\ \citenamefont
  {Tze}(1975)}]{Ezawa:1975ua}%
  \BibitemOpen
  \bibfield  {author} {\bibinfo {author} {\bibfnamefont {Z.~F.}\ \bibnamefont
  {Ezawa}}\ and\ \bibinfo {author} {\bibfnamefont {H.~C.}\ \bibnamefont
  {Tze}},\ }\href {https://doi.org/10.1016/0550-3213(75)90582-9} {\bibfield
  {journal} {\bibinfo  {journal} {Nucl. Phys. B}\ }\textbf {\bibinfo {volume}
  {96}},\ \bibinfo {pages} {264} (\bibinfo {year} {1975})}\BibitemShut
  {NoStop}%
\bibitem [{\citenamefont {Vachaspati}\ and\ \citenamefont
  {Vilenkin}(1985)}]{PhysRevD.31.3052}%
  \BibitemOpen
  \bibfield  {author} {\bibinfo {author} {\bibfnamefont {T.}~\bibnamefont
  {Vachaspati}}\ and\ \bibinfo {author} {\bibfnamefont {A.}~\bibnamefont
  {Vilenkin}},\ }\href {https://doi.org/10.1103/PhysRevD.31.3052} {\bibfield
  {journal} {\bibinfo  {journal} {Phys. Rev. D}\ }\textbf {\bibinfo {volume}
  {31}},\ \bibinfo {pages} {3052} (\bibinfo {year} {1985})}\BibitemShut
  {NoStop}%
\bibitem [{\citenamefont {Sakellariadou}(1990)}]{Sakellariadou1990}%
  \BibitemOpen
  \bibfield  {author} {\bibinfo {author} {\bibfnamefont {M.}~\bibnamefont
  {Sakellariadou}},\ }\href {https://doi.org/10.1103/PhysRevD.42.354}
  {\bibfield  {journal} {\bibinfo  {journal} {Phys. Rev. D}\ }\textbf {\bibinfo
  {volume} {42}},\ \bibinfo {pages} {354} (\bibinfo {year} {1990})}\BibitemShut
  {NoStop}%
\bibitem [{\citenamefont {Battye}\ and\ \citenamefont
  {Shellard}(1994)}]{Battye:1993jv}%
  \BibitemOpen
  \bibfield  {author} {\bibinfo {author} {\bibfnamefont {R.~A.}\ \bibnamefont
  {Battye}}\ and\ \bibinfo {author} {\bibfnamefont {E.~P.~S.}\ \bibnamefont
  {Shellard}},\ }\href {https://doi.org/10.1016/0550-3213(94)90573-8}
  {\bibfield  {journal} {\bibinfo  {journal} {Nucl. Phys. B}\ }\textbf
  {\bibinfo {volume} {423}},\ \bibinfo {pages} {260} (\bibinfo {year}
  {1994})},\ \Eprint {https://arxiv.org/abs/astro-ph/9311017}
  {arXiv:astro-ph/9311017} \BibitemShut {NoStop}%
\bibitem [{\citenamefont {Vilenkin}\ and\ \citenamefont
  {Shellard}(1994)}]{vilenkin1994cosmic}%
  \BibitemOpen
  \bibfield  {author} {\bibinfo {author} {\bibfnamefont {A.}~\bibnamefont
  {Vilenkin}}\ and\ \bibinfo {author} {\bibfnamefont {E.}~\bibnamefont
  {Shellard}},\ }\href {https://books.google.pl/books?id=eW4bB\_LAthEC} {\emph
  {\bibinfo {title} {Cosmic Strings and Other Topological Defects}}},\
  Cambridge Monographs on Mathematical Physics\ (\bibinfo  {publisher}
  {Cambridge University Press},\ \bibinfo {year} {1994})\BibitemShut {NoStop}%
\bibitem [{\citenamefont {Nash}\ and\ \citenamefont
  {Sen}(1988)}]{nash1988topology}%
  \BibitemOpen
  \bibfield  {author} {\bibinfo {author} {\bibfnamefont {C.}~\bibnamefont
  {Nash}}\ and\ \bibinfo {author} {\bibfnamefont {S.}~\bibnamefont {Sen}},\
  }\href
  {http://www.amazon.com/Topology-Geometry-Physicists-Charles-Nash/dp/0125140819/ref=sr_1_1?ie=UTF8&s=books&qid=1263990064&sr=1-1}
  {\emph {\bibinfo {title} {Topology and Geometry for Physicists}}}\ (\bibinfo
  {publisher} {Academic Press},\ \bibinfo {year} {1988})\BibitemShut {NoStop}%
\bibitem [{\citenamefont {Acquaviva}\ \emph {et~al.}(2021)\citenamefont
  {Acquaviva}, \citenamefont {Iorio},\ and\ \citenamefont
  {Smaldone}}]{Acquaviva:2020cjx}%
  \BibitemOpen
  \bibfield  {author} {\bibinfo {author} {\bibfnamefont {G.}~\bibnamefont
  {Acquaviva}}, \bibinfo {author} {\bibfnamefont {A.}~\bibnamefont {Iorio}},\
  and\ \bibinfo {author} {\bibfnamefont {L.}~\bibnamefont {Smaldone}},\ }\href
  {https://doi.org/10.1016/j.aop.2021.168641} {\bibfield  {journal} {\bibinfo
  {journal} {Annals Phys.}\ }\textbf {\bibinfo {volume} {434}},\ \bibinfo
  {pages} {168641} (\bibinfo {year} {2021})}\BibitemShut {NoStop}%
\end{thebibliography}%

\end{document}